\documentclass[twocolumn,
prb,
amsmath,
amssymb,
superscriptaddress]{revtex4-2}

\usepackage{graphicx}
\usepackage{hyperref}
\usepackage{siunitx}
\usepackage{comment}
\usepackage[version=4]{mhchem}
\usepackage[utf8]{inputenc}
\usepackage{csquotes}
\usepackage{float}
\usepackage{tikz}
\usetikzlibrary{arrows.meta}

\providecommand{\vect}[1]{{\boldsymbol{#1}}}
\newcommand{\Ms}{M_{\mathrm{s}}}

\begin{document}
\title{Spacetime magnetic hopfions:
from internal excitations and braiding of skyrmions}

\author{R.\ Knapman}
\affiliation{Institute of Physics, Johannes Gutenberg University Mainz, 55128 Mainz, Germany}
\affiliation{Faculty of Physics, University of Duisburg-Essen, 47057 Duisburg, Germany}
\affiliation{Center for Nanointegration Duisburg-Essen (CENIDE), University of Duisburg-Essen, 47057 Duisburg, Germany}

\author{T.\ Tausendpfund}
\affiliation{Institute of Physics, Johannes Gutenberg University Mainz, 55128 Mainz, Germany}

\author{S.\ A.\ D\'{i}az}
\affiliation{Faculty of Physics, University of Duisburg-Essen, 47057 Duisburg, Germany}
\affiliation{Department of Physics, University of Konstanz, 78457 Konstanz, Germany}

\author{K.\ Everschor-Sitte}
\affiliation{Faculty of Physics, University of Duisburg-Essen, 47057 Duisburg, Germany}
\affiliation{Center for Nanointegration Duisburg-Essen (CENIDE), University of Duisburg-Essen, 47057 Duisburg, Germany}

\begin{abstract}
Spatial topology endows topological solitons, such as skyrmions and hopfions, with fascinating dynamics.
However, the temporal dimension has so far provided a passive stage on which topological solitons evolve.
Here we construct spacetime magnetic hopfions: magnetic textures in two spatial dimensions that when excited by a time-periodic drive develop spacetime topology.
We uncover two complementary construction routes using skyrmions by braiding their center of mass position and by controlling their internal low-energy excitations.
Spacetime magnetic hopfions can be realized in nanopatterned grids to braid skyrmions and in frustrated magnets under an applied AC electric field.
Their topological invariant, the spacetime Hopf index, can be tuned by the applied electric field as demonstrated by our collective coordinate modeling and micromagnetic simulations.
The principles we have introduced to actively control spacetime topology are not limited to magnetic solitons, opening avenues to explore spacetime topology of general order parameters and fields.
\end{abstract}

\maketitle

\section{Introduction}

Topological solitons are robust localized structures actively studied by disciplines ranging from cosmology~\cite{Kibble1976,Rubakov2017} and particle physics~\cite{Skyrme1962,Rajaraman1982,Ryder1996,MantonSutcliffe} to optics~\cite{Dirac1931,Sugic2021,Shen2021,Zdagkas2022} and condensed matter~\cite{Mermin1979,Heeger1988,Altland2010}. 
Despite the differences across the many systems where they form, topological solitons exhibit enhanced stability which is due to the nontrivial topology of their structure in space.
In contrast, the temporal dimension does not play a role in defining soliton topology.

Topological solitons in magnets include skyrmions and hopfions~\cite{Fert2017, Everschor-Sitte2018, Back2020,Masell2021,Gobel2021,Tokura2021}.
The skyrmion topology is determined by the number of times the magnetization wraps the unit sphere. 
On the other hand, the hopfion structure rests on curves in space where the magnetization points in the same direction, its preimages~\cite{Hopf1931, Faddeev1997, Sutcliffe2007,Sutcliffe2017, Rybakov2022}. 
How often any two preimages wind around each other is counted by the Hopf index~\cite{Hopf1931}, the invariant that characterizes hopfion topology.
Although skyrmions and hopfions exhibit fascinating dynamics, it is rooted exclusively in their spatial topology, a property of their spatial structure that they carry as they evolve in time.

However, robust structures in time, including spacetime crystals in magnets, have been recently reported~\cite{Zhang2017a,Khemani2019,Trager2021,delSer2021,Bhowmick2022}. 
These reports motivate the question on the existence of topological solitons with time involved fundamentally and inextricably.
Here we turn to magnetic textures in two spatial dimensions and their time evolution to construct topological magnetic solitons in spacetime. Since their nontrivial spacetime topology is characterized by an extension of the Hopf index, we name them spacetime magnetic hopfions.

\begin{figure*}
    \centering
    \begin{tikzpicture}
        \node {\includegraphics[width=\linewidth]{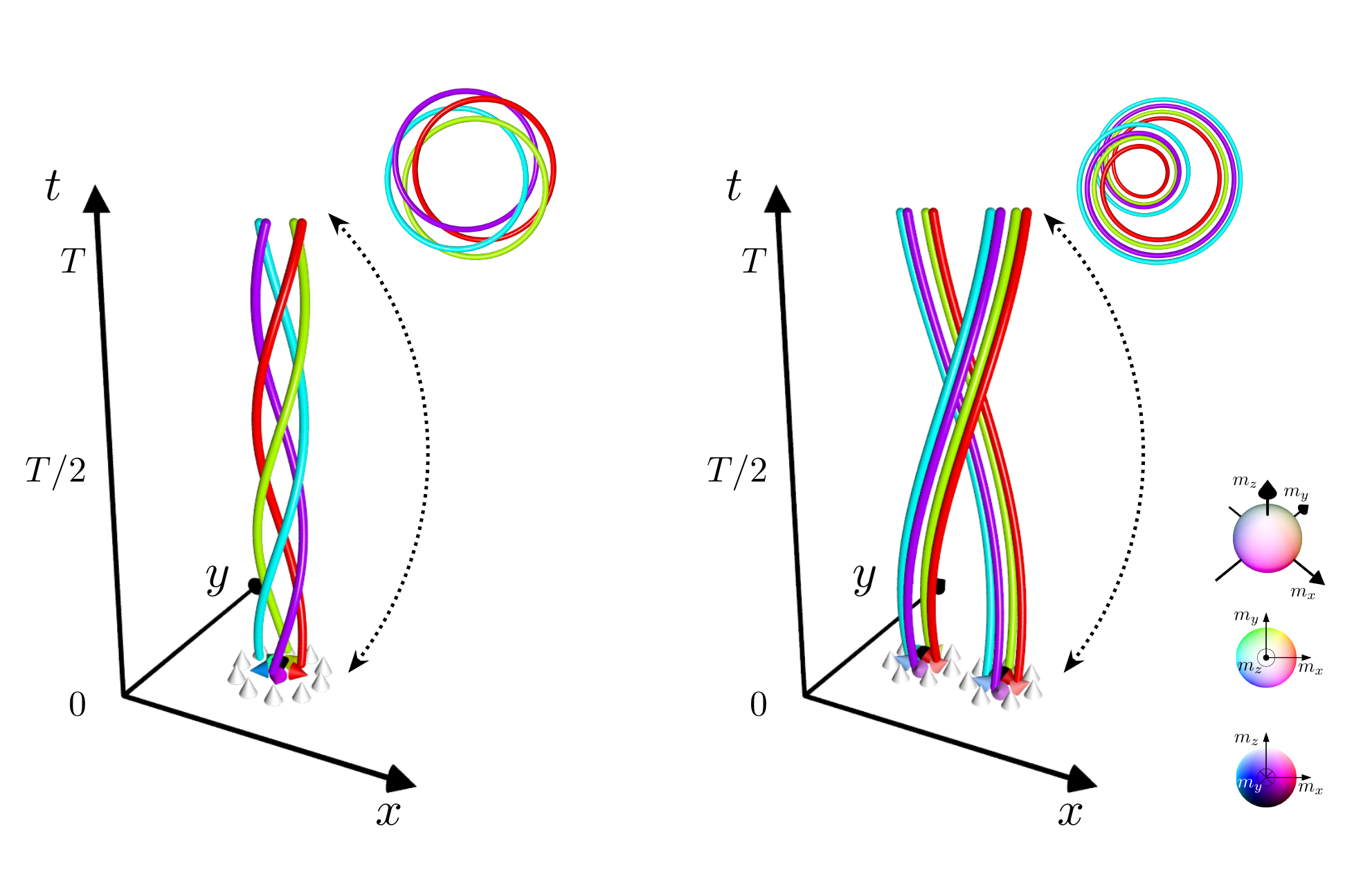}};
        \node[scale=1.25] at (-9, 4.5) {a};
        \node[scale=1.25] at (0, 4.5) {b};
    \end{tikzpicture}
    \caption{
    Spacetime magnetic hopfions---two construction routes based on skyrmion textures evolving periodically in time.
    a) Helicity rotation: the helicity of a skyrmion rotates by $2\pi$.
    b) Skyrmion braiding: two skyrmions swap positions.
    Preimages (colored curves) trace four selected magnetization directions throughout one period $T$.
    Identifying the magnetic textures, and their corresponding preimages, at $t=0$ and $t=T$, reveals their nontrivial spacetime topology.
    The preimages of the identified textures (insets) are linked, both corresponding to an $H = +1$ spacetime Hopf index.}
    \label{fig:IntroFigure}
\end{figure*}

\section{Results and Discussion}

\subsection{Spacetime topology construction}

To construct spacetime magnetic hopfions we exploit the topology of regular hopfions.

A hopfion in three spatial dimensions results from the following sequence of operations on a single skyrmion in two spatial dimensions.
Stack copies of the skyrmion to form a straight skyrmion tube.
Keeping one end of the tube fixed, twist the top end by an angle of 2$\pi$ about the tube axis which modifies the internal structure of the skyrmion along the tube.
The ends of the twisted tube are identical. 
Therefore, they can be identified by bending the tube and gluing together the matching ends. 
A three-dimensional hopfion has been formed.
If the tube had been twisted $n$ times by 2$\pi$, the resulting hopfion would have had a Hopf index equal to $nQ$, where $Q$ is the topological charge of the starting skyrmion.

In the preceding abstract construction, the skyrmion tube was built by stacking skyrmion copies.
It might as well have been generated from a skyrmion in two spatial dimensions whose texture is twisted as it evolved in time (see Fig.~\ref{fig:IntroFigure}a).
Moreover, such time evolution must be periodic so the temporal dimension, as implied by the gluing of the tube ends, has the topology of a circle ($S^1$). 
Equivalently, we require a skyrmion whose helicity, namely, the angle between the radial and the in-plane magnetization direction, 
periodically rotates in time.
Therefore, to construct a spacetime magnetic hopfion we need a two-dimensional skyrmion with a time-periodic rotating helicity.

There is an alternative route to construct spacetime magnetic hopfions. 
Instead of one, it requires (at least) two spatially separated skyrmions. 
By braiding the skyrmions' center of mass positions we induce time-periodicity.
Note that over the span of one period the skyrmions swap positions (see Fig.~\ref{fig:IntroFigure}b).
After identifying the swapped and initial skyrmion configurations, the resulting spacetime structure is topologically equivalent to a $2\pi$-twisted skyrmion tube with its ends identified, and therefore to a spacetime magnetic hopfion.

Spacetime magnetic hopfions constructed from time-periodic, two-dimensional skyrmionic textures, as described above, have nontrivial spacetime topology. 
The topological invariant suitable for their characterization is the spacetime 
Hopf index    
\begin{align}
    H = -\frac{1}{(8\pi)^2} \int \mathrm{d} t \, \int \mathrm{d}^2 r \, \vect{A}_{\mathrm{e}} \cdot \vect{B}_{\mathrm{e}} \,,
\end{align}
where $\vect{A}_{\mathrm{e}}$ is the emergent gauge field satisfying $\vect{B}_{\mathrm{e}} = \nabla \times \vect{A}_{\mathrm{e}}$, $\vect{B}_{\mathrm{e}}$ is the emergent magnetic field given by $\vect{B}_{\mathrm{e}, i} = \epsilon_{ijk} \vect{m} \cdot (\partial_j \vect{m} \times \partial_k \vect{m})$, with $\vect{m}$ the normalized magnetization, and $i, j, k \in \{x, y, t\}$. 
While the spatial integration is over the entire $x$-$y$ plane, time is integrated over one period $T$. 

In the following, we discuss how to concretely realize both routes to construct spacetime magnetic hopfions: by controlling the internal low-energy excitations (rotating helicity) of a single skyrmion, and by braiding the center of mass positions of multiple skyrmions.

\subsection{Route 1: single skyrmion internal excitations}

One route to construct spacetime magnetic hopfions relies on making the skyrmion helicity $\eta$ time-periodic, i.e., $\eta (t) = \eta (t+T)$, where $T$ is the period.
However, the skyrmion-hosting magnetic material typically fixes the helicity at a value that is energetically favorable.
Therefore, we first need to inject energy into the system, which will activate the internal excitations of the skyrmion and perhaps provide a way of controlling the helicity. 
These internal excitations might introduce unwanted multiply-periodic time dependence or, even worse, destabilize and eventually destroy the skyrmion itself.
A more targeted strategy is to identify a class of skyrmion-hosting materials where the helicity is among the lowest-energy excitations. 
Even better, the lowest-energy one with vanishing excitation energy: a Goldstone mode.

\begin{figure*}[t!]
	\centering
        \begin{tikzpicture}
    	\node {\includegraphics[width=\linewidth]{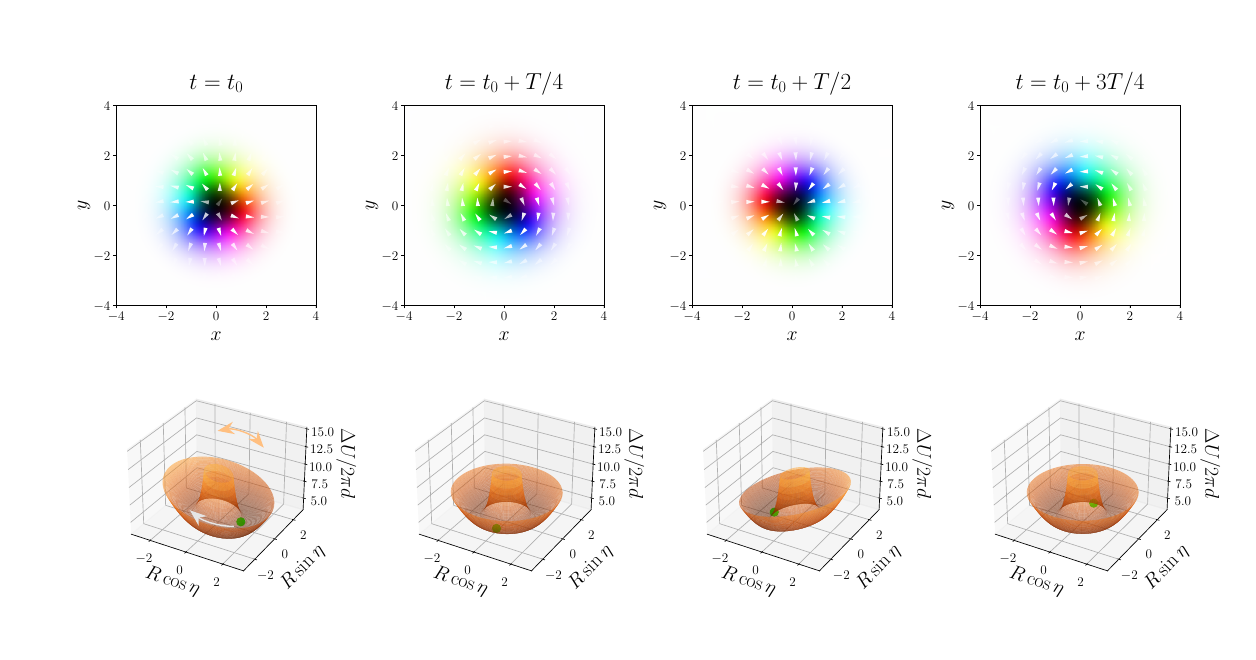}};
            \node[scale=1.25] at (-8, 3.5) {a};
            \node[scale=1.25] at (-8, -1) {b};
        \end{tikzpicture}
	\caption{
 Time evolution of a skyrmion in a frustrated magnet driven by an AC electric field: activating helicity rotation.
 The skyrmion dynamics reaches a time-periodic steady state synchronizing to the driving electric field's period $T$.
 a) Skyrmion texture snapshots, at $T/4$ intervals, from micromagnetic simulations confirm the $T$-periodic evolution.
 b) Energy landscape snapshots as functions of the skyrmion collective coordinates, helicity $\eta$ and radius $R$, shown in the $R\cos\eta - R\sin\eta$ plane with $\Delta U = U  + \int \mathrm{d}^3 r \, \vect{B} \cdot \hat{\vect{z}}$ 
 and sample thickness $d$.
  During one period, the energy landscape rocks back and forth about the $R\cos\eta = 0$ line as indicated by the orange arrow. 
 The green dot represents the skyrmion collective coordinates time evolution. 
 The direction of the helicity rotation (grey curved arrow) is uniquely defined by the driving electric field. 
 Activating the $T$-periodic helicity rotation forms a spacetime magnetic hopfion.
 Parameters: $E_0=0.20$, $\omega=2\pi/T=1.18$, $t_0$ is deep in the steady state and $\eta(t_0) = 0$ (mod $2\pi$); more details in Methods.
  }
	\label{fig:HelicityRotation}
\end{figure*}

The skyrmion helicity is known to be a Goldstone mode in frustrated magnets~\cite{Leonov2015,Lin2016,Zhang2017,Yao2020}.
Therefore, in the absence of energy injection and dissipation, energy conservation enforces the skyrmion size to remain constant while allowing the helicity to rotate at a constant angular frequency.
Another crucial property of frustrated magnets is that noncollinear magnetic textures, such as skyrmions, induce electric polarization $\vect{P}$~\cite{Katsura2005,Yao2020,Psaroudaki2021}. 
An applied electric field $\vect{E}$ couples to the electric polarization 
as $\vect{E} \cdot \vect{P}$.
This coupling provides a convenient, energy-efficient way to use a time-dependent applied electric field to control the magnetic skyrmion and hence, as we show below, its helicity.

Aiming to describe a broad class of frustrated magnets, we use the following rescaled micromagnetic energy functional
\begin{equation}
\label{eq:EnergyFunc}
    U = \int \mathrm{d}^3 r  \Bigl[ -\frac{1}{2} (\nabla \vect{m})^2 + \frac{1}{2} (\nabla^2 \vect{m})^2 
    - \vect{B} \cdot \vect{m} - \vect{E} \cdot \vect{P} \Bigr],
\end{equation}
with adimensionalized external magnetic and electric fields $\vect{B}$ and $\vect{E}$ (see Methods). 
In the following, we set $\vect{B}=\hat{\vect{z}}$ and assume a thin film sample geometry.
The interplay of the quadratic and quartic exchange terms captures the details of the competing microscopic exchange interactions. 
The competing exchange terms give rise to a length scale, which, in the presence of a moderate out-of-plane magnetic field, stabilizes skyrmions~\cite{Lin2016}.
Noncollinear magnetic textures, e.g.\ skyrmions, induce the electric polarization
\begin{equation}
	\label{eq:Polarization}
	\vect{P} = (\nabla \cdot \vect{m}) \vect{m} - (\vect{m} \cdot \nabla) \vect{m} \,,
\end{equation}
which couples to the applied electric field.
The coupling has the form of an effective Dzyaloshinskii-Moriya interaction (DMI):  $\vect E \cdot \vect P = D^j_{ik}(\vect E)\, m_i \partial_j m_k$ with $D^{j}_{ik}(\vect E) = \epsilon_{ikm}\epsilon_{mlj} E_l$.
Thus, the electric field allows the direct manipulation of the strength and sign of the effective DMI.
For an electric field applied along the direction perpendicular to the thin film, the $\vect{E} \cdot \vect{P}$ coupling breaks rotational symmetry and provides control over the skyrmion helicity.
Thus, this coupling allows an AC electric field to drive the frustrated magnet's magnetization dynamics and activate the helicity mode.

To model the electric-field-driven magnetization dynamics of a skyrmion we utilize two complementary approaches: micromagnetic simulations and collective coordinates.
Our micromagnetic simulations employ the above micromagnetic energy functional, Eq.~\eqref{eq:EnergyFunc}, together with the Landau-Lifshitz-Gilbert (LLG) equation (see Methods).
In the collective coordinate modeling we focus on the two most relevant degrees of freedom: the skyrmion helicity $\eta$ and, as a proxy for its canonically conjugate variable, the radius $R$.
We derive their generalized Thiele equations, two coupled first-order nonlinear differential equations, which govern the joint skyrmion helicity and radius-driven dynamics, and we solve them numerically (see Methods).
Both modeling approaches allow us to include energy injection from the driving AC electric field and energy dissipation from Gilbert damping, as expected in magnetic materials.
Below, we show how to construct a spacetime magnetic hopfion by controlling the skyrmion dynamics driven by an alternating electric field.

An applied AC electric field $\vect{E}(t) = E_0 \cos(\omega t) \hat{\vect{z}}$ drives the skyrmion dynamics.
In this case, the electric field coupling induces a cosine helicity dependence, i.e.\ $\vect{E} \cdot \vect{P}\propto \cos\eta$.
After the initial transient, the skyrmion dynamics reaches a time-periodic steady state with the same period $T = 2\pi/\omega$ of the driving AC electric field.
The top row in Fig.~\ref{fig:HelicityRotation} shows selected snapshots of the skyrmion texture during such a steady state obtained from micromagnetic simulations.
The helicity executes a full rotation during one period of the applied AC electric field signaling the formation of a spacetime magnetic hopfion.

Further understanding of the driven helicity rotation follows from the micromagnetic energy, Eq.~\eqref{eq:EnergyFunc}, depicted in the bottom row of Fig.~\ref{fig:HelicityRotation} as a function of the collective coordinates, i.e.\ the skyrmion radius $R$ and the helicity $\eta$.
We find it convenient to plot the energy on the $R\cos\eta-R\sin\eta$ plane
as it reflects the rotational invariance of the system in the absence of the electric field.
The energy increases as the skyrmion shrinks below its equilibrium radius due to the quartic exchange interaction. 
In the presented collective coordinate model this yields an unphysical \enquote{horn}-shaped divergence at $R = 0$, which is an artifact of our skyrmion profile approximation for small radii.
As the skyrmion expands beyond its equilibrium radius, the energy also increases due to the Zeeman term: $- \vect{B} \cdot \vect{m}$.
Thus, there results an energy minimum \enquote{trough.}
During one period the energy landscape rocks back and forth about the $R\cos\eta = 0$ line, such that the global energy minimum is at $\eta = \pi$ ($\eta = 0$) for $E_0 > 0$ ($E_0 < 0$). The amount of tilting is proportional to the electric field strength.
The skyrmion’s collective coordinates temporal evolution is represented by the green dot.
The direction of the helicity rotation is uniquely defined by the applied drive, 
a feature that can be understood from the Thiele equations (see Methods for details)
\begin{subequations}
\begin{align}
\dot{R} &= c_1 \alpha+ E_0 \cos(\omega t)\,(c_2 + c_3 \alpha) +\mathcal{O}(\alpha^2), \\
\dot{\eta}&= \dot{\eta}_0(R)+ E_0 \cos(\omega t)\,(c_4 + c_5 \alpha)  +\mathcal{O}(\alpha^2),
\end{align}
\end{subequations}
where $\alpha$ is the damping parameter and the functions $c_1, \dots, c_5$ depend on the collective coordinates. $\dot{\eta}_0(R)$ is the helicity rotation speed in the limit of $E_0=0$ and zero damping.
 In this limit, the magnetic energy $U$ does not depend explicitly on the helicity (it is a Goldstone mode), and attains its minimum at the critical radius $R^*$. 
 Therefore, the Thiele equations decouple and reduce to $\dot{R} = 0$ while $\dot{\eta}=\dot{\eta}_0(R)$ becomes only a function of $R$ and swaps sign at a critical skyrmion radius $R^*$. For $R < R^*$, the helicity evolves clockwise (decreases, $\dot{\eta}_0 < 0$), while for $R > R^*$, it evolves counterclockwise (increases, $\dot{\eta}_0 > 0$), see Supplementary Information.
In the presence of small damping and for $E_0=0$ the radius decreases to $R^*$ and the speed of the helicity rotation decreases to zero. The electric field drives the helicity rotation actively and compensates for the damping.  

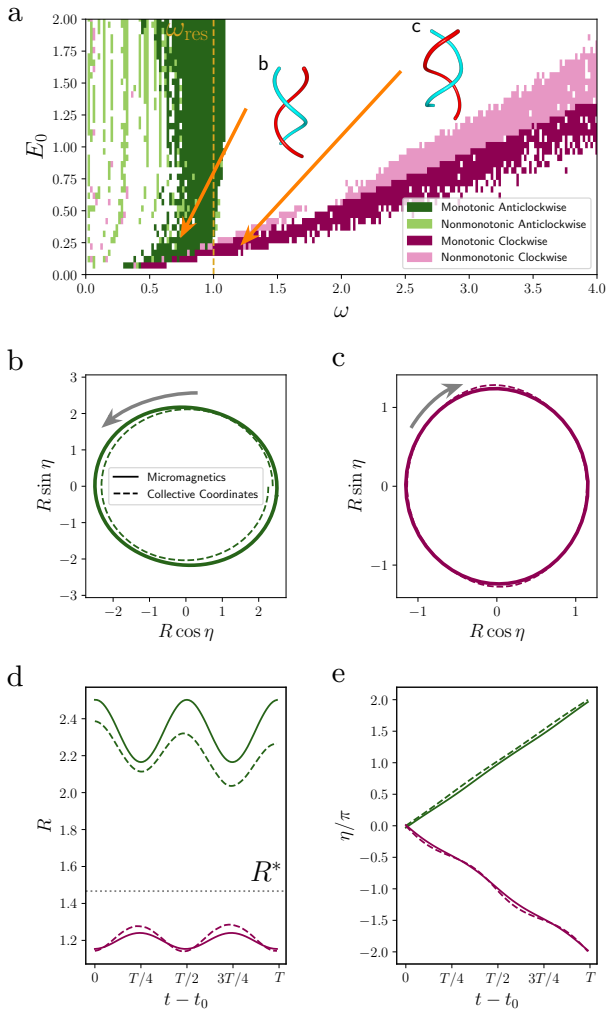
\begin{figure}[t]
    \centering
    \begin{tikzpicture}
        \node {\includegraphics[width=0.98\linewidth]{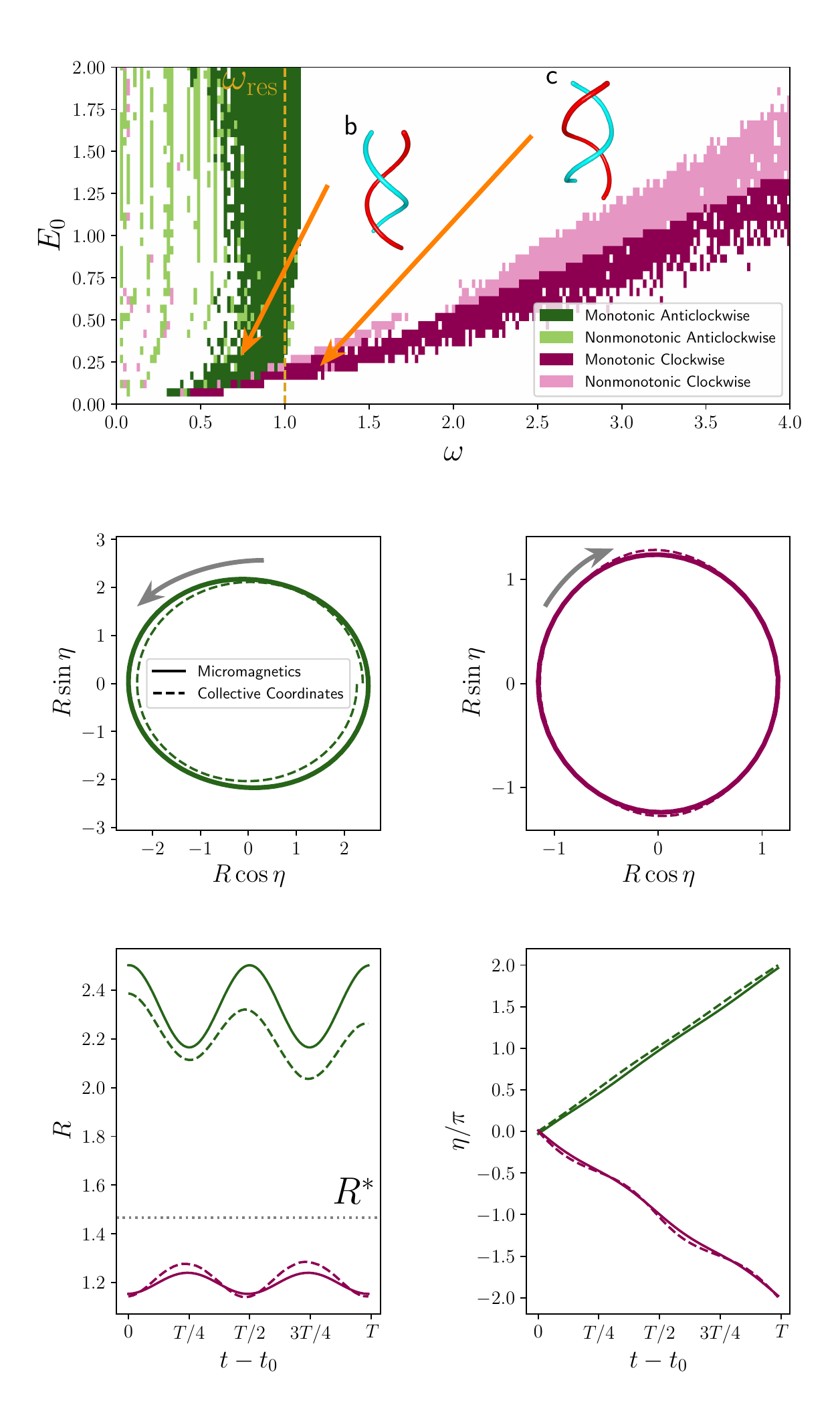}};
        \node[scale=1.25] at (-4, 6.5) {a};
        \node[scale=1.25] at (-4, 1.9) {b};
        \node[scale=1.25] at (0.3, 1.9) {c};
        \node[scale=1.25] at (-4, -2.3) {d};
        \node[scale=1.25] at (0.3, -2.3) {e};
    \end{tikzpicture}
    \caption{
    Roadmap to construct spacetime magnetic hopfions by electric-field-driven helicity rotation.
    a) Phase diagram of the long-term helicity dynamics, built from collective coordinate modeling with Gilbert damping $\alpha = 0.01$, of a skyrmion driven by an AC electric field with amplitude $E_0$ and frequency $\omega$.
    The helicity rotates clockwise (counterclockwise) in the pink (green) regions leading to nonzero spacetime Hopf indices as indicated by the linked preimages 
    of two selected points: b (c) with $H=-1$ ($H=+1$).
    b), c) Trajectories in the $R\cos\eta - R\sin\eta$ plane of the selected points from a) with parameters  
    $E_0 = 0.25$, $\omega = 0.72$ in b); and $E_0 = 0.20$, $\omega = 1.18$ in c).
    The trajectories are elliptical due to the skyrmion radius oscillations (breathing) shown in d).
    When the radius is larger (smaller) than $R^*$, the helicity increases (decreases) e), i.e., it rotates counterclockwise (clockwise).}
    \label{fig:PhaseDiagram}
\end{figure}

To investigate the conditions for helicity rotation required for a spacetime magnetic hopfion, we solved the Thiele equations for a range of electric field amplitudes $E_{0}$ and frequencies $\omega$.
From the steady-state solutions we built the phase diagram in Fig.~\ref{fig:PhaseDiagram}a.
In dark pink (dark green), we show solutions for which the skyrmion helicity rotates monotonically clockwise (counterclockwise), and in light pink (light green) the solutions for which the skyrmion helicity performs a non-monotonic but overall clockwise (counterclockwise) rotation.
Below the pink region, the electric field amplitude is too low to induce a helicity rotation, while between the pink and dark green regions, the dynamics of the system is complex, and the helicity rotations do not synchronize with the electric field oscillation frequency. Helicity rotation solutions, the colored regions, cluster around two \enquote{lobes.}

In the leftmost lobe, dominated by counterclockwise helicity evolution (dark and light green regions), the Thiele equations predict the formation of spacetime magnetic hopfions with spacetime Hopf index $H = -1$.
They form in the vicinity of the Kittel resonance frequency $\omega_{\rm{res}}$. 
Since our model does not include dipolar interactions, $\omega_{\rm{res}}\propto B_z$; in dimensionless units, having set $B_z= 1$, $\omega_{\rm{res}} = 1$.
Around the Kittel resonance frequency and for high $E_{0}$ energy injection supersedes dissipation.
This imbalance is reflected in  Thiele equations solutions with unphysically diverging skyrmion radii, signaling that the collective coordinate model is no longer adequate in this resonance region.
While micromagnetic simulations of skyrmions in this region do not show diverging radii, they reveal spin wave emission and more complex skyrmion distortions.
Even though the skyrmion helicity remains time-periodic, the resonance effects tend to disrupt the time periodicity of the skyrmion profile and thus the spacetime magnetic hopfion formation in this region.
Nevertheless, in the rest of the leftmost lobe, the Thiele equations and micromagnetic simulations are in good agreement, as exemplified by the (solid and dashed) dark green trajectories in Figs.~\ref{fig:PhaseDiagram}b, d, and e.
We conclude that in the leftmost lobe a low electric field amplitude $E_{0}$ and $\omega$ sufficiently far from $\omega_{\mathrm{res}}$ ensure a moderate energy injection-dissipation imbalance and thus guarantee the formation of spacetime magnetic hopfions with $H = -1$.

The rightmost lobe comprises only clockwise helicity evolution (dark and light pink regions) corresponding to spacetime magnetic hopfions with spacetime Hopf index $H = +1$.
Here, resonance effects are absent, so the Thiele equations solutions and micromagnetic simulations are in good agreement, see e.g.\ the (solid and dashed) dark pink trajectories in Figs.~\ref{fig:PhaseDiagram}
c,d, and e. Animations of the motion of the collective coordinates and the energy landscape, as well as the micromagnetics simulations, for both selected points shown in Fig.~\ref{fig:PhaseDiagram}, are shown in the supplementary movies.
The energy injection rate is balanced by or slightly below the dissipation rate.
Therefore, due to the balanced injection-dissipation ratio (see Supplementary Information), the absence of resonant effects, and a sizeable range of electric field amplitude and frequency, the rightmost lobe is an attractive region for the experimental realization of spacetime magnetic hopfions with $H =+1$.

A noteworthy feature of the right lobe is that the regions where the helicity evolution is monotonic (dark pink) and non-monotonic (pink) are separated by a straight line.
This straight line coincides with Thiele equations solutions for which the maximum radius attained during a helicity rotation cycle equals $R^*$ (see Supplementary Information). 
Therefore, for values of $E_{0}$ and $\omega$ in the pink region, the skyrmion radius crosses $R = R^*$ causing the helicity to reverse direction, thus becoming non-monotonic.

The trajectories in the $R\cos\eta-R\sin\eta$ plane of two selected points on the phase diagram, depicted in Figs.~\ref{fig:PhaseDiagram}b,c, have an elliptical shape.
They reflect the coupled collective coordinate dynamics; helicity rotations are accompanied by skyrmion breathing, i.e.\ radial oscillations.
When the radius is larger (smaller) than $R^*$, Fig.~\ref{fig:PhaseDiagram}d, the helicity increases (decreases), Fig.~\ref{fig:PhaseDiagram}e, which corresponds to counterclockwise (clockwise) rotations. 
For trajectories belonging to the phase diagram's leftmost (green) lobe, as in Fig.~\ref{fig:PhaseDiagram}b, the radius attains its maxima at $\eta = 0, \pi$ and minima at $\eta = \pi/2, 3\pi/2$.
The opposite holds for trajectories from the rightmost (pink) lobe, as in Fig.~\ref{fig:PhaseDiagram}c, whose radius maxima are at $\eta = \pi/2, 3\pi/2$ and minima at $\eta = 0, \pi$.
Since the helicity dependence of the micromagnetic energy comes solely from the coupling $\vect{E} \cdot \vect{P}\propto \cos\eta$, the rocking of the energy landscape (see Fig.~\ref{fig:HelicityRotation}b) gives trajectories with $R>R^*$ access to larger values of $R$ along the $R\cos\eta$ than the $R\sin\eta$ axis.
The situation reverses for trajectories with $R<R^*$.
Therefore, the difference between these two elliptical trajectories boils down to the size of $R$ relative to $R^*$.

The above results confirm that spacetime magnetic hopfions are not rare and isolated, on the contrary, they can be constructed over a large range of frequencies and amplitudes of the driving electric field.
Furthermore, the phase diagram in Fig.~\ref{fig:PhaseDiagram} provides a guide to tune between spacetime magnetic hopfions with opposite spacetime Hopf indices.

\subsection{Route 2: multiple skyrmion braiding}

An alternative construction route of spacetime magnetic hopfions follows from multiple skyrmion braiding.
Braiding skyrmions requires exchanging their center of mass positions (see Fig.~\ref{fig:IntroFigure}b).
Controlling the position of skyrmions is crucial to many data storage and computing applications~\cite{Fert2013,Zhang2015a}.
For instance, braiding skyrmions is at the core of a recent proposal for a topological quantum computing platform~\cite{Nothhelfer2022}.
Therefore, the manipulation of skyrmion positions has rapidly developed and it is currently technologically advanced~\cite{Fert2017,Li2023}.

Control of skyrmion positions has been achieved in chiral, multilayer, and frustrated magnets. The breadth of mechanism to set individual skyrmions into motion include electric currents~\cite{Jonietz2010,Komineas2015,Woo2018,Zhang2020a,Xia2019,Hou2020}, tilted magnetic fields~\cite{Moon2015}, magnetic field gradients~\cite{Zhang2018}, temperature gradients~\cite{Raimondo2022}, and circularly polarized laser illumination~\cite{Tengdin2022}.
The above mechanisms can be modified to also induce rotational motion of skyrmion crystals~\cite{Everschor2012}.
Models to simulate the above mechanisms of driven skyrmion dynamics are already available.

Instead of simulating the dynamics of a particular model, here we focus on characterizing the spacetime topology arising from skyrmion braiding.
We designed the magnetic texture of two skyrmions whose positions orbit each other (counterclockwise and clockwise) periodically over time.
The period is defined by the time it takes the skyrmions to swap their positions.
We then numerically computed their spacetime Hopf index.
We find that for a counterclockwise swap (as in Fig.~\ref{fig:IntroFigure}b) the Hopf index is $H = +1$, while for a clockwise swap $H = -1$.

It is conceptually straightforward to generalize the above construction to accommodate an arbitrary number of skyrmions.
They could be assembled into skyrmion crystals, skyrmion bags~\cite{Tang2021,Kind2021}, or engineered arrangements in nanopatterned substrates~\cite{Nothhelfer2022,Nothhelferpatent2019}.
Braiding multiple skyrmions would result in spacetime magnetic hopfions with high spacetime Hopf indices.

\subsection{Discussion}

The two spacetime magnetic hopfion construction routes we have presented can be experimentally realized in various material platforms.
Skyrmion braiding (construction route 2) requires controlling the skyrmion positions; this can be achieved by exploiting 
the experimentally reported skyrmion motion in the metallic chiral magnet FeGe~\cite{Yu2012,Yu2020}, the insulating chiral magnet Cu$_2$OSeO$_3$~\cite{Seki2012b,White2014,Zhang2018c}, the van der Waals ferromagnet Fe$_3$GeTe$_2$~\cite{Ding2020}, and in magnetic multilayers~\cite{Jiang2016,Woo2016}.
The rotation of the skyrmion helicity (construction route 1)
is expected in frustrated magnets such as the skyrmion-hosting materials Gd$_2$PdSi$_3$~\cite{Kurumaji2019}, and GdRu$_2$Si$_2$~\cite{Khanh2020}.
Alternatively, we expect it to appear in skyrmion-hosting systems with tunable or weak DMI and for dynamically stabilized skyrmions where the helicity can be actively tuned~\cite{McKeever2019, Zhou2015}.

Our spacetime magnetic hopfion constructions rely on magnetic textures  whose spatial extension is two-dimensional; good approximations to this limit are the commonly fabricated thin films.
Therefore, our modeling and simulations of frustrated magnets assume a thin film sample geometry, where (nonlocal) dipolar interactions, in the zero thickness limit, are well approximated by a (local) uniaxial magnetic anisotropy~\cite{Winter1961,Rohart2013}.
Aiming at a simple and universal description, we have neglected anisotropies and dipolar interactions.

In frustrated magnets and multiferroics~\cite{Mostovoy2006,Cheong2007,Seki2012} it is possible that the
the noncollinear magnetic texture of a single skyrmion induces electric polarization whose divergence is proportional to the bound electric charge density.
While an axially symmetric skyrmion induces no net charge, it produces a spatial rearrangement of bound charge highly localized at the skyrmion core.
The oscillating radius of a breathing skyrmion would make its attached bound charge oscillate and hence radiate electromagnetic waves.
Consequently, measuring this electromagnetic radiation could serve as a mechanism of indirect skyrmion breathing detection.

Direct detection of spacetime magnetic hopfions calls for measuring and tracking over time skyrmion magnetic textures; suitable for this purpose are Lorentz transmission electron microscopy (TEM)~\cite{Yu2010} and electron holography~\cite{Park2014}.
If the stability and time-periodic evolution of skyrmions in a particular system has already been established, it could be sufficient to just measure the helicity (construction route 1) or position (construction route 2).
The skyrmion helicity can be extracted from Lorentz TEM measurements~\cite{Shibata2013}, and in the case of crystals or disordered ensembles of skyrmions, it can be measured using circularly polarized resonant elastic x-ray scattering~\cite{Zhang2018b}. 
Skyrmion position measurements have been reported using scanning transmission X-ray microscopy (STXM)~\cite{Woo2016,Woo2017,Litzius2017}, magneto-optical Kerr effect (MOKE) microscopy~\cite{Jiang2016,Jiang2017,Velez2022,Quessab2022}, and magnetic force microscopy (MFM)~\cite{Casiraghi2019,Meng2019,Raju2019,Zhang2022}.

\section{Conclusion}

We have constructed spacetime magnetic hopfions.
They are magnetic solitons whose spacetime structure, characterized by the spacetime Hopf index, carries nontrivial spacetime topology.
We have shown two complementary construction routes that leverage the spatial topology of two-dimensional skyrmionic textures and a time-periodic drive: by skyrmion helicity rotation and by braiding skyrmions.
Readily available experimental platforms to realize these construction routes of spacetime magnetic hopfions are frustrated magnets under an applied AC electric field, and chiral magnets or magnetic multilayers driven by an applied AC magnetic field.
The principles we have used to construct spacetime magnetic hopfions can be applied beyond hopfions and magnetism.
Other systems, where topological solitons also emerge, might include liquid crystals, superfluids, superconductors, and electromagnetic fields.
We envisage the time-periodic manipulation of solitons with low-dimensional spatial topology as a versatile method to develop high-dimensional spacetime topology.
This method can become a tool for the active control of spacetime topology of general order parameters and fields.

\section{Methods}

The spacetime magnetic hopfions and, in particular, the electric-field-driven magnetization dynamics of a skyrmion considered in this work are modeled by the Landau-Lifshitz-Gilbert (LLG) equation~\cite{Landau1935,Gilbert2004}
\begin{equation}
	\label{eq:LLG}
	\frac{\partial \vect{m}}{\partial t} = -\gamma \vect{m} \times \vect{B}_{\mathrm{eff}} + \alpha \vect{m} \times \frac{\partial \vect{m}}{\partial t} \, ,
\end{equation}
where $\gamma$ is the electron gyromagnetic ratio, $\alpha$ is the Gilbert damping constant, and $\vect{B}_{\mathrm{eff}} = -\Ms^{-1} \delta U / \delta \vect{m}$ is the effective magnetic field with the saturation magnetization $\Ms$.

The dimensionful energy functional is given by
\begin{equation}
    	U =\int \Big[-\frac{I_1}{2}(\nabla \vect{m})^{2} + \frac{I_2}{2}\left(\nabla^{2} \vect{m}\right)^{2} 
 - \Ms \vect{B} \cdot \vect{m} - \vect{E} \cdot \vect{P} \Big]\, \mathrm{d}V,
 \label{eq:Udim}
\end{equation}
where $I_1$ and $I_2$ are strengths of the quadratic and quartic exchange interactions, respectively, and $\vect{B}$ ($\vect{E}$) is the externally applied magnetic (electric) field. 
Noncollinear magnetic textures induce the electric polarization
\begin{equation}
	\vect{P} = P_E a [(\nabla \cdot \vect{m}) \vect{m} - (\vect{m} \cdot \nabla) \vect{m}],
 \label{eq:Pdim}
\end{equation}
where $a$ is the lattice constant and $P_E$ is the polarization density, which couples to the applied electric field. Rescaling physical quantities as summarized in Table~\ref{tab:units} leaves the theory with three parameters, the rescaled dimensionless externally applied fields and the Gilbert damping parameter, $\alpha = 0.01$ in this work. Moreover, Eqs.~\eqref{eq:Udim} and \eqref{eq:Pdim}  
translate into Eqs.~\eqref{eq:EnergyFunc} and \eqref{eq:Polarization} in the main text. We have solved the skyrmion dynamics subjected to a periodic electric field using two complementary approaches: micromagnetic simulations and collective coordinates.

\begin{table}
	\centering
	\begin{tabular}{|c|c|c|c|}
		\hline
		Quantity & Characteristic & Definition & SI Unit \\
		\hline
		$\ell$ & Length & $\sqrt{I_2 / I_1}$ & \SI{}{\meter} \\
		$\tau$ & Time & $\Ms I_2 / \gamma I_1^2$ & \SI{}{\second} \\
		$\mathcal{U}$ & Energy & $\sqrt{I_1 I_2}$ & \SI{}{\joule} \\
		$\mathcal{B}$ & 
  Magnetic field
  & $I_1^2 / \Ms I_2$ & \SI{}{\tesla} \\
		$\mathcal{E}$ & Electric field & $(\gamma / \Ms^2) \sqrt{I_1^7 / I_2^3}$ & \SI{}{\volt\per\meter} \\
		$\mathcal{P}$ & Polarization & $(\Ms^2 / \gamma) \sqrt{I_2 / I_1^3}$ & \SI{}{\coulomb\per\square\meter} \\
		\hline
	\end{tabular}
	\caption{The physical constants by which our dimensionless quantities must be multiplied by to obtain their values in SI units.}
	\label{tab:units}
\end{table}

\subsection{Micromagnetic Simulations}

For our micromagnetic simulations, we use the open-source micromagnetics simulation package \textsc{MuMax3}~\cite{Vansteenkiste2014} with self-written extensions for the 
quartic exchange interaction and the electric field term. We neglect the effects of the demagnetizing field in our simulations for consistency with the collective coordinate modeling. 

We consider a single-cell thin-film sample with $N = 128$ cells in $x$ and $y$ directions.
Our system is discretized in cubic cells with edge length $\Delta = 0.3\sqrt{I_2/I_1}$.
To suppress the reflection of spin waves from the thin film edges, we set $\alpha = 1$ over an area bordering the edges and extending up to $5$ simulation cells into the sample.

We initialize our simulations with a skyrmion texture embedded in a ferromagnetic background and relax the system using the \texttt{Minimize()} function.

For the results shown in Figs.~\ref{fig:HelicityRotation} and~\ref{fig:PhaseDiagram} we used $\alpha = 0.01$, $I_1 = 10^{-12}\SI{}{\joule\per\meter}$, $I_2 =10^{-29} \SI{}{\joule\meter}$ and $\Ms = 10^6\SI{}{\ampere\per\meter}$. For these parameter choices we obtain, for example, a magnetic field of $\SI{100}{\milli\tesla}=1.0 \, I_1^2 / \Ms I_2$ and a typical length scale of $\SI{31.6}{\nano\meter} \approx 10 \, \sqrt{I_2 / I_1} $) for a dimensionless 
magnetic field of \num{1.0} and a dimensionless length of \num{10}.
By micromagnetic simulations, we have also confirmed that we get the same results when we rescale the parameters according to Table~\ref{tab:units}.

To calculate the radii of the skyrmions, we extract the contour for which $m_z = 0$ using the \textsc{scikit-image} library~\cite{skimage}, and calculate the mean of the displacements from the center of the contour. To compute the helicities, we extract the average helicity of the points on the $m_z = 0$ contour.

\subsection{Collective Coordinates}
\label{sec:CollectiveCoordinates}

In the collective coordinate modeling~\cite{Tretiakov2008,Clarke2008} of the skyrmion, we employed rescaled units. We focus on the temporal dynamics of the two most relevant degrees of freedom: the helicity $\eta(t)$ and, as a proxy for its canonically conjugate variable, the radius $R(t)$~\cite{McKeever2019}.

Representing the magnetization by spherical angles, $\vect{m}= (\cos\phi \sin\theta, \sin\phi \sin\theta, \cos\theta)$, a rotationally invariant skyrmion texture centered at the origin of the $xy$ plane is parametrized by its in-plane angle $\phi = \phi (\eta(t))$ and its profile function $\theta = \theta (R(t))$
which depend only on the skyrmion's helicity and radius, respectively.
We assume that the film thickness $d$ is sufficiently small such that magnetic textures can be considered uniform in the $z$-direction.

In this case, the effective equations of motions of the collective coordinate vector $\xi = (R,\eta)$ read $G_{ij}\dot{\xi}_j - \alpha \Gamma_{ij}\dot{\xi}_j + F_i = 0$ with the elements of gyrotropic $G$ and dissipative $\Gamma$ tensors, and generalized force $F$ given by
\begin{subequations}
\begin{align}
G_{ij} &= \int \vect{m} \cdot \left( \frac{\partial \vect{m}}{\partial \xi_i} \times \frac{\partial \vect{m}}{\partial \xi_j} \right)\, \mathrm{d}V ,\\
\Gamma_{ij} &=  \int \frac{\partial \vect{m}}{\partial \xi_i} \cdot \frac{\partial \vect{m}}{\partial \xi_j} \, \mathrm{d}V ,\\
F_i &= -\frac{\partial U}{\partial \xi_i}.
\end{align}
\end{subequations}
Only the generalized forces depend directly on the applied electric and magnetic fields.
For symmetry reasons, the antisymmetric gyrotropic tensor has only one independent non-vanishing tensor element, i.e.\ $G_{R\eta} = -G_{\eta R}$ and $G_{RR} = G_{\eta \eta}=0$, and the dissipative tensor is diagonal, i.e.\ $\Gamma_{\eta R} = \Gamma_{R \eta} = 0$. 
The effective equations of motion for the skyrmions radial's and helicity's temporal evolution become
\begin{equation}
    \begin{bmatrix} \dot{R} \\ \dot{\eta} \end{bmatrix} = \frac{1}{G_{R\eta}^2 + \alpha^2 \Gamma_{RR} \Gamma_{\eta\eta}} \begin{bmatrix} \alpha \Gamma_{\eta\eta} F_R + G_{R\eta} F_\eta \\ -G_{R\eta} F_R + \alpha \Gamma_{RR} F_\eta \end{bmatrix}.
    \label{eq:ThieleSystem}
\end{equation}
To solve Eq.~\eqref{eq:ThieleSystem} we use the following ansatz~\cite{Braun1994,Wang2018} for the skyrmion's in-plane angle $\phi$ and its profile function $\theta$ 
\begin{subequations}
\begin{align}
    \label{eq:PhiAnsatz}
    \phi &= m\psi + \eta(t), \\
    \label{eq:ThetaAnsatz}
    \theta &= 2 \arctan \left( \frac{\sinh(R(t)/w)}{\sinh(\rho/w)} \right),
\end{align}
\end{subequations}
where $\rho = \sqrt{x^2 + y^2}$ and $\psi = \arctan(y/x)$ are the radial and angular polar coordinates, respectively. The skyrmion vorticity $m$ is taken to be $1$ in this work, and $w$  characterizes the domain wall width of the skyrmion's profile.
We find that for a domain wall width of $w = 1.4$, this ansatz approximates well the skyrmion obtained by micromagnetic simulations for both cases: in the relaxed state when the electric field is absent, as well as for excited skyrmions upon stimulation with the electric field. 
Using this ansatz, the remaining tensor elements are independent of the skyrmion's helicity and depend only on its radius $R$ 
\begin{subequations}
\begin{align}
G_{R\eta} &= 2\pi d \int_0^\infty \rho \sin\theta \frac{\partial \theta}{\partial R} \, \mathrm{d} \rho,\\
\Gamma_{RR} &= 2\pi d \int_0^\infty \rho \left( \frac{\partial \theta}{\partial R} \right)^2 \, \mathrm{d} \rho ,\\
\Gamma_{\eta\eta}&= 2\pi d \int_0^\infty \rho \sin^2\theta \, \mathrm{d} \rho .
\end{align}
\end{subequations}
The generalized forces contain the information about the electric field breaking the rotational symmetry, which is why the electric field amplitude appears in combination with $\eta$:

\begin{subequations}
\begin{align}
F_R &=	E_0 \cos(\omega t) \cos\eta \, 2\pi d \int_0^\infty \Bigl(  \cos(2\theta) \frac{\partial \theta}{\partial R} + \rho \frac{\partial^2 \theta}{\partial R \partial \rho} \Bigr) \, \mathrm{d} \rho\notag\\
&\quad - \frac{\partial U_{\mathrm{ex}}}{\partial R} - B_z 2\pi d  \int_0^\infty\rho \sin\theta \frac{\partial \theta}{\partial R} \, \mathrm{d}\rho,\\
F_\eta &= - E_0 \cos(\omega t) \sin\eta \, 2 \pi d  \int_0^\infty \Bigl( \cos\theta \sin\theta +  \rho \frac{\partial \theta}{\partial \rho}\Bigr) \, \mathrm{d} \rho,
\end{align}
\end{subequations}
where $U_{\mathrm{ex}} = 2\pi d \int  [-\frac{1}{2} (\nabla \vect{m})^2 + \frac{1}{2} (\nabla^2 \vect{m})^2] \, \mathrm{d} \rho$ is the exchange interaction part of Eq.~\eqref{eq:EnergyFunc}.
Terms in $F_R$ that are independent of $E_0$ depend only on $R$.

We numerically integrate the collective coordinates up to a dimensionless time of \num{1000} with an adaptive time step size. 
In obtaining the phase diagram in Fig.~\ref{fig:PhaseDiagram}a, we took values in $0 \leq E_{0} \leq 2$ and $0 \leq \omega \leq 4$ using meshes with step sizes of \num{0.05} and \num{0.02}, respectively.
To obtain the number of helicity rotations per electric field cycle, we integrated the changes of the helicity angles and averaged them over all periods for $500 \leq t \leq 1000$, neglecting the initial transient dynamics.
Numerically, we classified a counterclockwise (clockwise) rotation as the average helicity rotation being greater than $0.9*2\pi$ (less than  $-0.9*2\pi$).  
To perform the many numerical time integrations of $R(t)$ and $\eta(t)$ efficiently, we use the \texttt{multiprocessing} Python package and the Radau IIA integrator supplied by \textsc{SciPy}~\cite{SciPy}. 
As initial conditions, we used the values of $R$ and $\eta$ that minimize the total energy Eq.~\eqref{eq:EnergyFunc} at $t=0$.

To avoid the computationally intensive evaluation of the integrals over the radial coordinate at each time integration step in $G_{R\eta}$, $\Gamma_{RR}$, $\Gamma_{\eta \eta}$, $F_R$, and $F_\eta$, we employed fit functions of the corresponding integrals in the range $0 \leq R \leq 10$.
We verified that the error due to fitting the integrals was negligible. 
In particular, we found good agreement for cases where  $R \lesssim 50$. 

We have confirmed that an externally applied magnetic field with $B_z = 2$ leads to qualitatively the same results shown in Fig.~\ref{fig:PhaseDiagram}a: with two lobes determining counterclockwise and clockwise helicity rotations and the resonance frequency being shifted to $\omega = 2$, as expected.

\section{Data Availability}

The corresponding collective coordinate evolution and micromagnetic simulation data, along with analysis scripts, are available 
upon reasonable request.

\section{Acknowledgements}

We thank Stefan Bl\"ugel, Nikolai Kiselev, and Volodymyr Kravchuk for fruitful discussions.
We acknowledge funding from the German Research Foundation (DFG) Project No.~320163632 (Emmy Noether), Project No.~403233384 (SPP2137 Skyrmionics), Project No.~505561633 (ANR/DFG TOROID).
R.~K.\ is supported by a scholarship from the Studienstiftung des deutschen Volkes. 

\bibliography{Bibliography.bib}

\begin{thebibliography}{92}%
\makeatletter
\providecommand \@ifxundefined [1]{%
 \@ifx{#1\undefined}
}%
\providecommand \@ifnum [1]{%
 \ifnum #1\expandafter \@firstoftwo
 \else \expandafter \@secondoftwo
 \fi
}%
\providecommand \@ifx [1]{%
 \ifx #1\expandafter \@firstoftwo
 \else \expandafter \@secondoftwo
 \fi
}%
\providecommand \natexlab [1]{#1}%
\providecommand \enquote  [1]{``#1''}%
\providecommand \bibnamefont  [1]{#1}%
\providecommand \bibfnamefont [1]{#1}%
\providecommand \citenamefont [1]{#1}%
\providecommand \href@noop [0]{\@secondoftwo}%
\providecommand \href [0]{\begingroup \@sanitize@url \@href}%
\providecommand \@href[1]{\@@startlink{#1}\@@href}%
\providecommand \@@href[1]{\endgroup#1\@@endlink}%
\providecommand \@sanitize@url [0]{\catcode `\\12\catcode `\$12\catcode `\&12\catcode `\#12\catcode `\^12\catcode `\_12\catcode `\%12\relax}%
\providecommand \@@startlink[1]{}%
\providecommand \@@endlink[0]{}%
\providecommand \url  [0]{\begingroup\@sanitize@url \@url }%
\providecommand \@url [1]{\endgroup\@href {#1}{\urlprefix }}%
\providecommand \urlprefix  [0]{URL }%
\providecommand \Eprint [0]{\href }%
\providecommand \doibase [0]{https://doi.org/}%
\providecommand \selectlanguage [0]{\@gobble}%
\providecommand \bibinfo  [0]{\@secondoftwo}%
\providecommand \bibfield  [0]{\@secondoftwo}%
\providecommand \translation [1]{[#1]}%
\providecommand \BibitemOpen [0]{}%
\providecommand \bibitemStop [0]{}%
\providecommand \bibitemNoStop [0]{.\EOS\space}%
\providecommand \EOS [0]{\spacefactor3000\relax}%
\providecommand \BibitemShut  [1]{\csname bibitem#1\endcsname}%
\let\auto@bib@innerbib\@empty
\bibitem [{\citenamefont {Kibble}(1976)}]{Kibble1976}%
  \BibitemOpen
  \bibfield  {author} {\bibinfo {author} {\bibfnamefont {T.~W.~B.}\ \bibnamefont {Kibble}},\ }\href {https://doi.org/10.1088/0305-4470/9/8/029} {\bibfield  {journal} {\bibinfo  {journal} {J. Phys. A: Math. Gen.}\ }\textbf {\bibinfo {volume} {9}},\ \bibinfo {pages} {1387} (\bibinfo {year} {1976})}\BibitemShut {NoStop}%
\bibitem [{\citenamefont {Rubakov}\ and\ \citenamefont {Gorbunov}(2017)}]{Rubakov2017}%
  \BibitemOpen
  \bibfield  {author} {\bibinfo {author} {\bibfnamefont {V.~A.}\ \bibnamefont {Rubakov}}\ and\ \bibinfo {author} {\bibfnamefont {D.~S.}\ \bibnamefont {Gorbunov}},\ }\bibinfo {title} {Topological defects and solitons in the universe},\ in\ \href {https://doi.org/10.1142/9789813220041\textunderscore0012} {\emph {\bibinfo {booktitle} {Introduction to the Theory of the Early Universe}}}\ (\bibinfo  {publisher} {World Scientific},\ \bibinfo {year} {2017})\ Chap.~\bibinfo {chapter} {12}, pp.\ \bibinfo {pages} {377--427},\ \bibinfo {edition} {2nd}\ ed.\BibitemShut {Stop}%
\bibitem [{\citenamefont {Skyrme}(1962)}]{Skyrme1962}%
  \BibitemOpen
  \bibfield  {author} {\bibinfo {author} {\bibfnamefont {T.}~\bibnamefont {Skyrme}},\ }\href {https://doi.org/10.1016/0029-5582(62)90775-7} {\bibfield  {journal} {\bibinfo  {journal} {Nucl. Phys.}\ }\textbf {\bibinfo {volume} {31}},\ \bibinfo {pages} {556} (\bibinfo {year} {1962})}\BibitemShut {NoStop}%
\bibitem [{\citenamefont {Wallace}(1983)}]{Rajaraman1982}%
  \BibitemOpen
  \bibfield  {author} {\bibinfo {author} {\bibfnamefont {D.~J.}\ \bibnamefont {Wallace}},\ }\href {https://doi.org/10.1088/0031-9112/34/1/043} {\bibfield  {journal} {\bibinfo  {journal} {Physics Bulletin}\ }\textbf {\bibinfo {volume} {34}},\ \bibinfo {pages} {29} (\bibinfo {year} {1983})}\BibitemShut {NoStop}%
\bibitem [{\citenamefont {Ryder}(1996)}]{Ryder1996}%
  \BibitemOpen
  \bibfield  {author} {\bibinfo {author} {\bibfnamefont {L.~H.}\ \bibnamefont {Ryder}},\ }\href {https://doi.org/10.1017/CBO9780511813900} {\emph {\bibinfo {title} {Quantum Field Theory}}},\ \bibinfo {edition} {2nd}\ ed.\ (\bibinfo  {publisher} {Cambridge University Press},\ \bibinfo {year} {1996})\BibitemShut {NoStop}%
\bibitem [{\citenamefont {Manton}\ and\ \citenamefont {Sutcliffe}(2004)}]{MantonSutcliffe}%
  \BibitemOpen
  \bibfield  {author} {\bibinfo {author} {\bibfnamefont {N.}~\bibnamefont {Manton}}\ and\ \bibinfo {author} {\bibfnamefont {P.}~\bibnamefont {Sutcliffe}},\ }\href {https://doi.org/10.1017/CBO9780511617034} {\emph {\bibinfo {title} {Topological solitons}}}\ (\bibinfo  {publisher} {Cambridge University Press},\ \bibinfo {year} {2004})\BibitemShut {NoStop}%
\bibitem [{\citenamefont {Dirac}(1931)}]{Dirac1931}%
  \BibitemOpen
  \bibfield  {author} {\bibinfo {author} {\bibfnamefont {P.~A.~M.}\ \bibnamefont {Dirac}},\ }\href {https://doi.org/10.1098/rspa.1931.0130} {\bibfield  {journal} {\bibinfo  {journal} {Proc. R. Soc. London A.}\ }\textbf {\bibinfo {volume} {133}},\ \bibinfo {pages} {60} (\bibinfo {year} {1931})}\BibitemShut {NoStop}%
\bibitem [{\citenamefont {Sugic}\ \emph {et~al.}(2021)\citenamefont {Sugic}, \citenamefont {Droop}, \citenamefont {Otte}, \citenamefont {Ehrmanntraut}, \citenamefont {Nori}, \citenamefont {Ruostekoski}, \citenamefont {Denz},\ and\ \citenamefont {Dennis}}]{Sugic2021}%
  \BibitemOpen
  \bibfield  {author} {\bibinfo {author} {\bibfnamefont {D.}~\bibnamefont {Sugic}}, \bibinfo {author} {\bibfnamefont {R.}~\bibnamefont {Droop}}, \bibinfo {author} {\bibfnamefont {E.}~\bibnamefont {Otte}}, \bibinfo {author} {\bibfnamefont {D.}~\bibnamefont {Ehrmanntraut}}, \bibinfo {author} {\bibfnamefont {F.}~\bibnamefont {Nori}}, \bibinfo {author} {\bibfnamefont {J.}~\bibnamefont {Ruostekoski}}, \bibinfo {author} {\bibfnamefont {C.}~\bibnamefont {Denz}},\ and\ \bibinfo {author} {\bibfnamefont {M.~R.}\ \bibnamefont {Dennis}},\ }\href {https://doi.org/10.1038/s41467-021-26171-5} {\bibfield  {journal} {\bibinfo  {journal} {Nat. Commun.}\ }\textbf {\bibinfo {volume} {12}},\ \bibinfo {pages} {6785} (\bibinfo {year} {2021})}\BibitemShut {NoStop}%
\bibitem [{\citenamefont {Shen}\ \emph {et~al.}(2021)\citenamefont {Shen}, \citenamefont {Hou}, \citenamefont {Papasimakis},\ and\ \citenamefont {Zheludev}}]{Shen2021}%
  \BibitemOpen
  \bibfield  {author} {\bibinfo {author} {\bibfnamefont {Y.}~\bibnamefont {Shen}}, \bibinfo {author} {\bibfnamefont {Y.}~\bibnamefont {Hou}}, \bibinfo {author} {\bibfnamefont {N.}~\bibnamefont {Papasimakis}},\ and\ \bibinfo {author} {\bibfnamefont {N.~I.}\ \bibnamefont {Zheludev}},\ }\href {https://doi.org/10.1038/s41467-021-26037-w} {\bibfield  {journal} {\bibinfo  {journal} {Nat. Commun.}\ }\textbf {\bibinfo {volume} {12}},\ \bibinfo {pages} {5891} (\bibinfo {year} {2021})}\BibitemShut {NoStop}%
\bibitem [{\citenamefont {Zdagkas}\ \emph {et~al.}(2022)\citenamefont {Zdagkas}, \citenamefont {McDonnell}, \citenamefont {Deng}, \citenamefont {Shen}, \citenamefont {Li}, \citenamefont {Ellenbogen}, \citenamefont {Papasimakis},\ and\ \citenamefont {Zheludev}}]{Zdagkas2022}%
  \BibitemOpen
  \bibfield  {author} {\bibinfo {author} {\bibfnamefont {A.}~\bibnamefont {Zdagkas}}, \bibinfo {author} {\bibfnamefont {C.}~\bibnamefont {McDonnell}}, \bibinfo {author} {\bibfnamefont {J.}~\bibnamefont {Deng}}, \bibinfo {author} {\bibfnamefont {Y.}~\bibnamefont {Shen}}, \bibinfo {author} {\bibfnamefont {G.}~\bibnamefont {Li}}, \bibinfo {author} {\bibfnamefont {T.}~\bibnamefont {Ellenbogen}}, \bibinfo {author} {\bibfnamefont {N.}~\bibnamefont {Papasimakis}},\ and\ \bibinfo {author} {\bibfnamefont {N.~I.}\ \bibnamefont {Zheludev}},\ }\href {https://doi.org/10.1038/s41566-022-01028-5} {\bibfield  {journal} {\bibinfo  {journal} {Nat. Photonics}\ }\textbf {\bibinfo {volume} {16}},\ \bibinfo {pages} {523} (\bibinfo {year} {2022})}\BibitemShut {NoStop}%
\bibitem [{\citenamefont {Mermin}(1979)}]{Mermin1979}%
  \BibitemOpen
  \bibfield  {author} {\bibinfo {author} {\bibfnamefont {N.~D.}\ \bibnamefont {Mermin}},\ }\href {https://doi.org/10.1103/RevModPhys.51.591} {\bibfield  {journal} {\bibinfo  {journal} {Rev. Mod. Phys.}\ }\textbf {\bibinfo {volume} {51}},\ \bibinfo {pages} {591} (\bibinfo {year} {1979})}\BibitemShut {NoStop}%
\bibitem [{\citenamefont {Heeger}\ \emph {et~al.}(1988)\citenamefont {Heeger}, \citenamefont {Kivelson}, \citenamefont {Schrieffer},\ and\ \citenamefont {Su}}]{Heeger1988}%
  \BibitemOpen
  \bibfield  {author} {\bibinfo {author} {\bibfnamefont {A.~J.}\ \bibnamefont {Heeger}}, \bibinfo {author} {\bibfnamefont {S.}~\bibnamefont {Kivelson}}, \bibinfo {author} {\bibfnamefont {J.~R.}\ \bibnamefont {Schrieffer}},\ and\ \bibinfo {author} {\bibfnamefont {W.~P.}\ \bibnamefont {Su}},\ }\href {https://doi.org/10.1103/RevModPhys.60.781} {\bibfield  {journal} {\bibinfo  {journal} {Rev. Mod. Phys.}\ }\textbf {\bibinfo {volume} {60}},\ \bibinfo {pages} {781} (\bibinfo {year} {1988})}\BibitemShut {NoStop}%
\bibitem [{\citenamefont {Altland}\ and\ \citenamefont {Simons}(2010)}]{Altland2010}%
  \BibitemOpen
  \bibfield  {author} {\bibinfo {author} {\bibfnamefont {A.}~\bibnamefont {Altland}}\ and\ \bibinfo {author} {\bibfnamefont {B.~D.}\ \bibnamefont {Simons}},\ }\href {https://doi.org/10.1017/CBO9780511789984} {\emph {\bibinfo {title} {Condensed Matter Field Theory}}},\ \bibinfo {edition} {2nd}\ ed.\ (\bibinfo  {publisher} {Cambridge University Press},\ \bibinfo {year} {2010})\BibitemShut {NoStop}%
\bibitem [{\citenamefont {Fert}\ \emph {et~al.}(2017)\citenamefont {Fert}, \citenamefont {Reyren},\ and\ \citenamefont {Cros}}]{Fert2017}%
  \BibitemOpen
  \bibfield  {author} {\bibinfo {author} {\bibfnamefont {A.}~\bibnamefont {Fert}}, \bibinfo {author} {\bibfnamefont {N.}~\bibnamefont {Reyren}},\ and\ \bibinfo {author} {\bibfnamefont {V.}~\bibnamefont {Cros}},\ }\href {https://doi.org/10.1038/natrevmats.2017.31} {\bibfield  {journal} {\bibinfo  {journal} {Nat. Rev. Mater.}\ }\textbf {\bibinfo {volume} {2}},\ \bibinfo {pages} {1} (\bibinfo {year} {2017})}\BibitemShut {NoStop}%
\bibitem [{\citenamefont {{Everschor-Sitte}}\ \emph {et~al.}(2018)\citenamefont {{Everschor-Sitte}}, \citenamefont {Masell}, \citenamefont {Reeve},\ and\ \citenamefont {Kl{\"a}ui}}]{Everschor-Sitte2018}%
  \BibitemOpen
  \bibfield  {author} {\bibinfo {author} {\bibfnamefont {K.}~\bibnamefont {{Everschor-Sitte}}}, \bibinfo {author} {\bibfnamefont {J.}~\bibnamefont {Masell}}, \bibinfo {author} {\bibfnamefont {R.~M.}\ \bibnamefont {Reeve}},\ and\ \bibinfo {author} {\bibfnamefont {M.}~\bibnamefont {Kl{\"a}ui}},\ }\href {https://doi.org/10.1063/1.5048972} {\bibfield  {journal} {\bibinfo  {journal} {J. Appl. Phys.}\ }\textbf {\bibinfo {volume} {124}},\ \bibinfo {pages} {240901} (\bibinfo {year} {2018})}\BibitemShut {NoStop}%
\bibitem [{\citenamefont {Back}\ \emph {et~al.}(2020)\citenamefont {Back}, \citenamefont {Cros}, \citenamefont {Ebert}, \citenamefont {{Everschor-Sitte}}, \citenamefont {Fert}, \citenamefont {Garst}, \citenamefont {Ma}, \citenamefont {Mankovsky}, \citenamefont {Monchesky}, \citenamefont {Mostovoy}, \citenamefont {Nagaosa}, \citenamefont {Parkin}, \citenamefont {Pfleiderer}, \citenamefont {Reyren}, \citenamefont {Rosch}, \citenamefont {Taguchi}, \citenamefont {Tokura}, \citenamefont {von Bergmann},\ and\ \citenamefont {Zang}}]{Back2020}%
  \BibitemOpen
  \bibfield  {author} {\bibinfo {author} {\bibfnamefont {C.}~\bibnamefont {Back}}, \bibinfo {author} {\bibfnamefont {V.}~\bibnamefont {Cros}}, \bibinfo {author} {\bibfnamefont {H.}~\bibnamefont {Ebert}}, \bibinfo {author} {\bibfnamefont {K.}~\bibnamefont {{Everschor-Sitte}}}, \bibinfo {author} {\bibfnamefont {A.}~\bibnamefont {Fert}}, \bibinfo {author} {\bibfnamefont {M.}~\bibnamefont {Garst}}, \bibinfo {author} {\bibfnamefont {T.}~\bibnamefont {Ma}}, \bibinfo {author} {\bibfnamefont {S.}~\bibnamefont {Mankovsky}}, \bibinfo {author} {\bibfnamefont {T.~L.}\ \bibnamefont {Monchesky}}, \bibinfo {author} {\bibfnamefont {M.}~\bibnamefont {Mostovoy}}, \bibinfo {author} {\bibfnamefont {N.}~\bibnamefont {Nagaosa}}, \bibinfo {author} {\bibfnamefont {S.~S.~P.}\ \bibnamefont {Parkin}}, \bibinfo {author} {\bibfnamefont {C.}~\bibnamefont {Pfleiderer}}, \bibinfo {author} {\bibfnamefont {N.}~\bibnamefont {Reyren}}, \bibinfo {author} {\bibfnamefont {A.}~\bibnamefont {Rosch}}, \bibinfo {author} {\bibfnamefont {Y.}~\bibnamefont {Taguchi}}, \bibinfo {author} {\bibfnamefont {Y.}~\bibnamefont {Tokura}}, \bibinfo {author} {\bibfnamefont {K.}~\bibnamefont {von Bergmann}},\ and\ \bibinfo {author} {\bibfnamefont {J.}~\bibnamefont {Zang}},\ }\href {https://doi.org/10.1088/1361-6463/ab8418} {\bibfield  {journal} {\bibinfo  {journal} {J. Phys. D: Appl. Phys.}\ }\textbf {\bibinfo {volume} {53}},\ \bibinfo {pages} {363001} (\bibinfo {year} {2020})}\BibitemShut {NoStop}%
\bibitem [{\citenamefont {Masell}\ and\ \citenamefont {{Everschor-Sitte}}(2021)}]{Masell2021}%
  \BibitemOpen
  \bibfield  {author} {\bibinfo {author} {\bibfnamefont {J.}~\bibnamefont {Masell}}\ and\ \bibinfo {author} {\bibfnamefont {K.}~\bibnamefont {{Everschor-Sitte}}},\ }in\ \href {https://doi.org/10.1007/978-3-030-62844-4_7} {\emph {\bibinfo {booktitle} {Chirality, Magnetism and Magnetoelectricity: {{Separate}} Phenomena and Joint Effects in Metamaterial Structures}}},\ \bibinfo {editor} {edited by\ \bibinfo {editor} {\bibfnamefont {E.}~\bibnamefont {Kamenetskii}}}\ (\bibinfo  {publisher} {{Springer International Publishing}},\ \bibinfo {address} {{Cham}},\ \bibinfo {year} {2021})\ pp.\ \bibinfo {pages} {147--181}\BibitemShut {NoStop}%
\bibitem [{\citenamefont {G{\"o}bel}\ \emph {et~al.}(2021)\citenamefont {G{\"o}bel}, \citenamefont {Mertig},\ and\ \citenamefont {Tretiakov}}]{Gobel2021}%
  \BibitemOpen
  \bibfield  {author} {\bibinfo {author} {\bibfnamefont {B.}~\bibnamefont {G{\"o}bel}}, \bibinfo {author} {\bibfnamefont {I.}~\bibnamefont {Mertig}},\ and\ \bibinfo {author} {\bibfnamefont {O.~A.}\ \bibnamefont {Tretiakov}},\ }\href {https://doi.org/10.1016/j.physrep.2020.10.001} {\bibfield  {journal} {\bibinfo  {journal} {Phys. Rep.}\ }\textbf {\bibinfo {volume} {895}},\ \bibinfo {pages} {1} (\bibinfo {year} {2021})}\BibitemShut {NoStop}%
\bibitem [{\citenamefont {Tokura}\ and\ \citenamefont {Kanazawa}(2021)}]{Tokura2021}%
  \BibitemOpen
  \bibfield  {author} {\bibinfo {author} {\bibfnamefont {Y.}~\bibnamefont {Tokura}}\ and\ \bibinfo {author} {\bibfnamefont {N.}~\bibnamefont {Kanazawa}},\ }\href {https://doi.org/10.1021/acs.chemrev.0c00297} {\bibfield  {journal} {\bibinfo  {journal} {Chem. Rev.}\ }\textbf {\bibinfo {volume} {121}},\ \bibinfo {pages} {2857} (\bibinfo {year} {2021})}\BibitemShut {NoStop}%
\bibitem [{\citenamefont {Hopf}(1931)}]{Hopf1931}%
  \BibitemOpen
  \bibfield  {author} {\bibinfo {author} {\bibfnamefont {H.}~\bibnamefont {Hopf}},\ }\href {https://doi.org/10.1007/978-3-662-25046-4_4} {\bibfield  {journal} {\bibinfo  {journal} {Math. Ann.}\ }\textbf {\bibinfo {volume} {104}},\ \bibinfo {pages} {637} (\bibinfo {year} {1931})}\BibitemShut {NoStop}%
\bibitem [{\citenamefont {Faddeev}\ and\ \citenamefont {Niemi}(1997)}]{Faddeev1997}%
  \BibitemOpen
  \bibfield  {author} {\bibinfo {author} {\bibfnamefont {L.}~\bibnamefont {Faddeev}}\ and\ \bibinfo {author} {\bibfnamefont {A.~J.}\ \bibnamefont {Niemi}},\ }\href {https://doi.org/10.1038/387058a0} {\bibfield  {journal} {\bibinfo  {journal} {Nature}\ }\textbf {\bibinfo {volume} {387}},\ \bibinfo {pages} {58} (\bibinfo {year} {1997})}\BibitemShut {NoStop}%
\bibitem [{\citenamefont {Sutcliffe}(2007)}]{Sutcliffe2007}%
  \BibitemOpen
  \bibfield  {author} {\bibinfo {author} {\bibfnamefont {P.}~\bibnamefont {Sutcliffe}},\ }\href {https://doi.org/10.1103/PhysRevB.76.184439} {\bibfield  {journal} {\bibinfo  {journal} {Phys. Rev. B}\ }\textbf {\bibinfo {volume} {76}},\ \bibinfo {pages} {184439} (\bibinfo {year} {2007})}\BibitemShut {NoStop}%
\bibitem [{\citenamefont {Sutcliffe}(2017)}]{Sutcliffe2017}%
  \BibitemOpen
  \bibfield  {author} {\bibinfo {author} {\bibfnamefont {P.}~\bibnamefont {Sutcliffe}},\ }\href {https://doi.org/10.1103/PhysRevLett.118.247203} {\bibfield  {journal} {\bibinfo  {journal} {Phys. Rev. Lett.}\ }\textbf {\bibinfo {volume} {118}},\ \bibinfo {pages} {247203} (\bibinfo {year} {2017})}\BibitemShut {NoStop}%
\bibitem [{\citenamefont {Rybakov}\ \emph {et~al.}(2022)\citenamefont {Rybakov}, \citenamefont {Kiselev}, \citenamefont {Borisov}, \citenamefont {D{\"o}ring}, \citenamefont {Melcher},\ and\ \citenamefont {Bl{\"u}gel}}]{Rybakov2022}%
  \BibitemOpen
  \bibfield  {author} {\bibinfo {author} {\bibfnamefont {F.~N.}\ \bibnamefont {Rybakov}}, \bibinfo {author} {\bibfnamefont {N.~S.}\ \bibnamefont {Kiselev}}, \bibinfo {author} {\bibfnamefont {A.~B.}\ \bibnamefont {Borisov}}, \bibinfo {author} {\bibfnamefont {L.}~\bibnamefont {D{\"o}ring}}, \bibinfo {author} {\bibfnamefont {C.}~\bibnamefont {Melcher}},\ and\ \bibinfo {author} {\bibfnamefont {S.}~\bibnamefont {Bl{\"u}gel}},\ }\href {https://doi.org/10.1063/5.0099942} {\bibfield  {journal} {\bibinfo  {journal} {APL Mater.}\ }\textbf {\bibinfo {volume} {10}},\ \bibinfo {pages} {111113} (\bibinfo {year} {2022})}\BibitemShut {NoStop}%
\bibitem [{\citenamefont {Zhang}\ \emph {et~al.}(2017{\natexlab{a}})\citenamefont {Zhang}, \citenamefont {Hess}, \citenamefont {Kyprianidis}, \citenamefont {Becker}, \citenamefont {Lee}, \citenamefont {Smith}, \citenamefont {Pagano}, \citenamefont {Potirniche}, \citenamefont {Potter}, \citenamefont {Vishwanath}, \citenamefont {Yao},\ and\ \citenamefont {Monroe}}]{Zhang2017a}%
  \BibitemOpen
  \bibfield  {author} {\bibinfo {author} {\bibfnamefont {J.}~\bibnamefont {Zhang}}, \bibinfo {author} {\bibfnamefont {P.~W.}\ \bibnamefont {Hess}}, \bibinfo {author} {\bibfnamefont {A.}~\bibnamefont {Kyprianidis}}, \bibinfo {author} {\bibfnamefont {P.}~\bibnamefont {Becker}}, \bibinfo {author} {\bibfnamefont {A.}~\bibnamefont {Lee}}, \bibinfo {author} {\bibfnamefont {J.}~\bibnamefont {Smith}}, \bibinfo {author} {\bibfnamefont {G.}~\bibnamefont {Pagano}}, \bibinfo {author} {\bibfnamefont {I.-D.}\ \bibnamefont {Potirniche}}, \bibinfo {author} {\bibfnamefont {A.~C.}\ \bibnamefont {Potter}}, \bibinfo {author} {\bibfnamefont {A.}~\bibnamefont {Vishwanath}}, \bibinfo {author} {\bibfnamefont {N.~Y.}\ \bibnamefont {Yao}},\ and\ \bibinfo {author} {\bibfnamefont {C.}~\bibnamefont {Monroe}},\ }\href {https://doi.org/10.1038/nature21413} {\bibfield  {journal} {\bibinfo  {journal} {Nature}\ }\textbf {\bibinfo {volume} {543}},\ \bibinfo {pages} {217} (\bibinfo {year} {2017}{\natexlab{a}})}\BibitemShut {NoStop}%
\bibitem [{\citenamefont {Khemani}\ \emph {et~al.}(2019)\citenamefont {Khemani}, \citenamefont {Moessner},\ and\ \citenamefont {Sondhi}}]{Khemani2019}%
  \BibitemOpen
  \bibfield  {author} {\bibinfo {author} {\bibfnamefont {V.}~\bibnamefont {Khemani}}, \bibinfo {author} {\bibfnamefont {R.}~\bibnamefont {Moessner}},\ and\ \bibinfo {author} {\bibfnamefont {S.~L.}\ \bibnamefont {Sondhi}},\ }\href {https://doi.org/10.48550/arXiv.1910.10745} {\bibinfo {title} {A brief history of time crystals}} (\bibinfo {year} {2019})\BibitemShut {NoStop}%
\bibitem [{\citenamefont {Tr\"ager}\ \emph {et~al.}(2021)\citenamefont {Tr\"ager}, \citenamefont {Gruszecki}, \citenamefont {Lisiecki}, \citenamefont {Gro\ss{}}, \citenamefont {F\"orster}, \citenamefont {Weigand}, \citenamefont {G\l{}owi\ifmmode~\acute{n}\else \'{n}\fi{}ski}, \citenamefont {Ku\ifmmode~\acute{s}\else \'{s}\fi{}wik}, \citenamefont {Dubowik}, \citenamefont {Sch\"utz}, \citenamefont {Krawczyk},\ and\ \citenamefont {Gr\"afe}}]{Trager2021}%
  \BibitemOpen
  \bibfield  {author} {\bibinfo {author} {\bibfnamefont {N.}~\bibnamefont {Tr\"ager}}, \bibinfo {author} {\bibfnamefont {P.}~\bibnamefont {Gruszecki}}, \bibinfo {author} {\bibfnamefont {F.}~\bibnamefont {Lisiecki}}, \bibinfo {author} {\bibfnamefont {F.}~\bibnamefont {Gro\ss{}}}, \bibinfo {author} {\bibfnamefont {J.}~\bibnamefont {F\"orster}}, \bibinfo {author} {\bibfnamefont {M.}~\bibnamefont {Weigand}}, \bibinfo {author} {\bibfnamefont {H.}~\bibnamefont {G\l{}owi\ifmmode~\acute{n}\else \'{n}\fi{}ski}}, \bibinfo {author} {\bibfnamefont {P.}~\bibnamefont {Ku\ifmmode~\acute{s}\else \'{s}\fi{}wik}}, \bibinfo {author} {\bibfnamefont {J.}~\bibnamefont {Dubowik}}, \bibinfo {author} {\bibfnamefont {G.}~\bibnamefont {Sch\"utz}}, \bibinfo {author} {\bibfnamefont {M.}~\bibnamefont {Krawczyk}},\ and\ \bibinfo {author} {\bibfnamefont {J.}~\bibnamefont {Gr\"afe}},\ }\href {https://doi.org/10.1103/PhysRevLett.126.057201} {\bibfield  {journal} {\bibinfo  {journal} {Phys. Rev. Lett.}\ }\textbf {\bibinfo {volume} {126}},\ \bibinfo {pages} {057201} (\bibinfo {year} {2021})}\BibitemShut {NoStop}%
\bibitem [{\citenamefont {{del Ser}}\ \emph {et~al.}(2021)\citenamefont {{del Ser}}, \citenamefont {Heinen},\ and\ \citenamefont {Rosch}}]{delSer2021}%
  \BibitemOpen
  \bibfield  {author} {\bibinfo {author} {\bibfnamefont {N.}~\bibnamefont {{del Ser}}}, \bibinfo {author} {\bibfnamefont {L.}~\bibnamefont {Heinen}},\ and\ \bibinfo {author} {\bibfnamefont {A.}~\bibnamefont {Rosch}},\ }\href {https://doi.org/10.21468/SciPostPhys.11.1.009} {\bibfield  {journal} {\bibinfo  {journal} {SciPost Phys.}\ }\textbf {\bibinfo {volume} {11}},\ \bibinfo {pages} {009} (\bibinfo {year} {2021})}\BibitemShut {NoStop}%
\bibitem [{\citenamefont {Bhowmick}\ \emph {et~al.}(2022)\citenamefont {Bhowmick}, \citenamefont {Sun}, \citenamefont {Yang},\ and\ \citenamefont {Sengupta}}]{Bhowmick2022}%
  \BibitemOpen
  \bibfield  {author} {\bibinfo {author} {\bibfnamefont {D.}~\bibnamefont {Bhowmick}}, \bibinfo {author} {\bibfnamefont {H.}~\bibnamefont {Sun}}, \bibinfo {author} {\bibfnamefont {B.}~\bibnamefont {Yang}},\ and\ \bibinfo {author} {\bibfnamefont {P.}~\bibnamefont {Sengupta}},\ }\href@noop {} {\  (\bibinfo {year} {2022})},\ \Eprint {https://arxiv.org/abs/2207.09077} {arXiv:2207.09077} \BibitemShut {NoStop}%
\bibitem [{\citenamefont {Leonov}\ and\ \citenamefont {Mostovoy}(2015)}]{Leonov2015}%
  \BibitemOpen
  \bibfield  {author} {\bibinfo {author} {\bibfnamefont {A.~O.}\ \bibnamefont {Leonov}}\ and\ \bibinfo {author} {\bibfnamefont {M.}~\bibnamefont {Mostovoy}},\ }\href {https://doi.org/10.1038/ncomms9275} {\bibfield  {journal} {\bibinfo  {journal} {Nat. Commun.}\ }\textbf {\bibinfo {volume} {6}},\ \bibinfo {pages} {8275} (\bibinfo {year} {2015})}\BibitemShut {NoStop}%
\bibitem [{\citenamefont {Lin}\ and\ \citenamefont {Hayami}(2016)}]{Lin2016}%
  \BibitemOpen
  \bibfield  {author} {\bibinfo {author} {\bibfnamefont {S.-Z.}\ \bibnamefont {Lin}}\ and\ \bibinfo {author} {\bibfnamefont {S.}~\bibnamefont {Hayami}},\ }\href {https://doi.org/10.1103/PhysRevB.93.064430} {\bibfield  {journal} {\bibinfo  {journal} {Phys. Rev. B}\ }\textbf {\bibinfo {volume} {93}},\ \bibinfo {pages} {064430} (\bibinfo {year} {2016})}\BibitemShut {NoStop}%
\bibitem [{\citenamefont {Zhang}\ \emph {et~al.}(2017{\natexlab{b}})\citenamefont {Zhang}, \citenamefont {Xia}, \citenamefont {Zhou}, \citenamefont {Liu}, \citenamefont {Zhang},\ and\ \citenamefont {Ezawa}}]{Zhang2017}%
  \BibitemOpen
  \bibfield  {author} {\bibinfo {author} {\bibfnamefont {X.}~\bibnamefont {Zhang}}, \bibinfo {author} {\bibfnamefont {J.}~\bibnamefont {Xia}}, \bibinfo {author} {\bibfnamefont {Y.}~\bibnamefont {Zhou}}, \bibinfo {author} {\bibfnamefont {X.}~\bibnamefont {Liu}}, \bibinfo {author} {\bibfnamefont {H.}~\bibnamefont {Zhang}},\ and\ \bibinfo {author} {\bibfnamefont {M.}~\bibnamefont {Ezawa}},\ }\href {https://doi.org/10.1038/s41467-017-01785-w} {\bibfield  {journal} {\bibinfo  {journal} {Nat. Commun.}\ }\textbf {\bibinfo {volume} {8}},\ \bibinfo {pages} {1717} (\bibinfo {year} {2017}{\natexlab{b}})}\BibitemShut {NoStop}%
\bibitem [{\citenamefont {Yao}\ \emph {et~al.}(2020)\citenamefont {Yao}, \citenamefont {Chen},\ and\ \citenamefont {Dong}}]{Yao2020}%
  \BibitemOpen
  \bibfield  {author} {\bibinfo {author} {\bibfnamefont {X.}~\bibnamefont {Yao}}, \bibinfo {author} {\bibfnamefont {J.}~\bibnamefont {Chen}},\ and\ \bibinfo {author} {\bibfnamefont {S.}~\bibnamefont {Dong}},\ }\href {https://doi.org/10.1088/1367-2630/aba1b3} {\bibfield  {journal} {\bibinfo  {journal} {New J. Phys.}\ }\textbf {\bibinfo {volume} {22}},\ \bibinfo {pages} {083032} (\bibinfo {year} {2020})}\BibitemShut {NoStop}%
\bibitem [{\citenamefont {Katsura}\ \emph {et~al.}(2005)\citenamefont {Katsura}, \citenamefont {Nagaosa},\ and\ \citenamefont {Balatsky}}]{Katsura2005}%
  \BibitemOpen
  \bibfield  {author} {\bibinfo {author} {\bibfnamefont {H.}~\bibnamefont {Katsura}}, \bibinfo {author} {\bibfnamefont {N.}~\bibnamefont {Nagaosa}},\ and\ \bibinfo {author} {\bibfnamefont {A.~V.}\ \bibnamefont {Balatsky}},\ }\href {https://doi.org/10.1103/PhysRevLett.95.057205} {\bibfield  {journal} {\bibinfo  {journal} {Phys. Rev. Lett.}\ }\textbf {\bibinfo {volume} {95}},\ \bibinfo {pages} {057205} (\bibinfo {year} {2005})}\BibitemShut {NoStop}%
\bibitem [{\citenamefont {Psaroudaki}\ and\ \citenamefont {Panagopoulos}(2021)}]{Psaroudaki2021}%
  \BibitemOpen
  \bibfield  {author} {\bibinfo {author} {\bibfnamefont {C.}~\bibnamefont {Psaroudaki}}\ and\ \bibinfo {author} {\bibfnamefont {C.}~\bibnamefont {Panagopoulos}},\ }\href {https://doi.org/10.1103/PhysRevLett.127.067201} {\bibfield  {journal} {\bibinfo  {journal} {Phys. Rev. Lett.}\ }\textbf {\bibinfo {volume} {127}},\ \bibinfo {pages} {067201} (\bibinfo {year} {2021})}\BibitemShut {NoStop}%
\bibitem [{\citenamefont {Fert}\ \emph {et~al.}(2013)\citenamefont {Fert}, \citenamefont {Cros},\ and\ \citenamefont {Sampaio}}]{Fert2013}%
  \BibitemOpen
  \bibfield  {author} {\bibinfo {author} {\bibfnamefont {A.}~\bibnamefont {Fert}}, \bibinfo {author} {\bibfnamefont {V.}~\bibnamefont {Cros}},\ and\ \bibinfo {author} {\bibfnamefont {J.}~\bibnamefont {Sampaio}},\ }\href {https://doi.org/10.1038/nnano.2013.29} {\bibfield  {journal} {\bibinfo  {journal} {Nat. Nanotechnol.}\ }\textbf {\bibinfo {volume} {8}},\ \bibinfo {pages} {152} (\bibinfo {year} {2013})}\BibitemShut {NoStop}%
\bibitem [{\citenamefont {Zhang}\ \emph {et~al.}(2015)\citenamefont {Zhang}, \citenamefont {Ezawa},\ and\ \citenamefont {Zhou}}]{Zhang2015a}%
  \BibitemOpen
  \bibfield  {author} {\bibinfo {author} {\bibfnamefont {X.}~\bibnamefont {Zhang}}, \bibinfo {author} {\bibfnamefont {M.}~\bibnamefont {Ezawa}},\ and\ \bibinfo {author} {\bibfnamefont {Y.}~\bibnamefont {Zhou}},\ }\href {https://doi.org/10.1038/srep09400} {\bibfield  {journal} {\bibinfo  {journal} {Sci. Rep.}\ }\textbf {\bibinfo {volume} {5}},\ \bibinfo {pages} {9400} (\bibinfo {year} {2015})}\BibitemShut {NoStop}%
\bibitem [{\citenamefont {Nothhelfer}\ \emph {et~al.}(2022)\citenamefont {Nothhelfer}, \citenamefont {D\'{\i}az}, \citenamefont {Kessler}, \citenamefont {Meng}, \citenamefont {Rizzi}, \citenamefont {Hals},\ and\ \citenamefont {Everschor-Sitte}}]{Nothhelfer2022}%
  \BibitemOpen
  \bibfield  {author} {\bibinfo {author} {\bibfnamefont {J.}~\bibnamefont {Nothhelfer}}, \bibinfo {author} {\bibfnamefont {S.~A.}\ \bibnamefont {D\'{\i}az}}, \bibinfo {author} {\bibfnamefont {S.}~\bibnamefont {Kessler}}, \bibinfo {author} {\bibfnamefont {T.}~\bibnamefont {Meng}}, \bibinfo {author} {\bibfnamefont {M.}~\bibnamefont {Rizzi}}, \bibinfo {author} {\bibfnamefont {K.~M.~D.}\ \bibnamefont {Hals}},\ and\ \bibinfo {author} {\bibfnamefont {K.}~\bibnamefont {Everschor-Sitte}},\ }\href {https://doi.org/10.1103/PhysRevB.105.224509} {\bibfield  {journal} {\bibinfo  {journal} {Phys. Rev. B}\ }\textbf {\bibinfo {volume} {105}},\ \bibinfo {pages} {224509} (\bibinfo {year} {2022})}\BibitemShut {NoStop}%
\bibitem [{\citenamefont {Li}\ \emph {et~al.}(2023)\citenamefont {Li}, \citenamefont {Wang},\ and\ \citenamefont {Rasing}}]{Li2023}%
  \BibitemOpen
  \bibfield  {author} {\bibinfo {author} {\bibfnamefont {S.}~\bibnamefont {Li}}, \bibinfo {author} {\bibfnamefont {X.}~\bibnamefont {Wang}},\ and\ \bibinfo {author} {\bibfnamefont {T.}~\bibnamefont {Rasing}},\ }\href {https://doi.org/10.1002/idm2.12072} {\bibfield  {journal} {\bibinfo  {journal} {Interdiscip. Mater.}\ }\textbf {\bibinfo {volume} {2}},\ \bibinfo {pages} {260} (\bibinfo {year} {2023})}\BibitemShut {NoStop}%
\bibitem [{\citenamefont {Jonietz}\ \emph {et~al.}(2010)\citenamefont {Jonietz}, \citenamefont {M{\"u}hlbauer}, \citenamefont {Pfleiderer}, \citenamefont {Neubauer}, \citenamefont {M{\"u}nzer}, \citenamefont {Bauer}, \citenamefont {Adams}, \citenamefont {Georgii}, \citenamefont {B{\"o}ni}, \citenamefont {Duine}, \citenamefont {Everschor}, \citenamefont {Garst},\ and\ \citenamefont {Rosch}}]{Jonietz2010}%
  \BibitemOpen
  \bibfield  {author} {\bibinfo {author} {\bibfnamefont {F.}~\bibnamefont {Jonietz}}, \bibinfo {author} {\bibfnamefont {S.}~\bibnamefont {M{\"u}hlbauer}}, \bibinfo {author} {\bibfnamefont {C.}~\bibnamefont {Pfleiderer}}, \bibinfo {author} {\bibfnamefont {A.}~\bibnamefont {Neubauer}}, \bibinfo {author} {\bibfnamefont {W.}~\bibnamefont {M{\"u}nzer}}, \bibinfo {author} {\bibfnamefont {A.}~\bibnamefont {Bauer}}, \bibinfo {author} {\bibfnamefont {T.}~\bibnamefont {Adams}}, \bibinfo {author} {\bibfnamefont {R.}~\bibnamefont {Georgii}}, \bibinfo {author} {\bibfnamefont {P.}~\bibnamefont {B{\"o}ni}}, \bibinfo {author} {\bibfnamefont {R.~A.}\ \bibnamefont {Duine}}, \bibinfo {author} {\bibfnamefont {K.}~\bibnamefont {Everschor}}, \bibinfo {author} {\bibfnamefont {M.}~\bibnamefont {Garst}},\ and\ \bibinfo {author} {\bibfnamefont {A.}~\bibnamefont {Rosch}},\ }\href {https://doi.org/10.1126/science.1195709} {\bibfield  {journal} {\bibinfo  {journal} {Science}\ }\textbf {\bibinfo {volume} {330}},\ \bibinfo {pages} {1648} (\bibinfo {year} {2010})}\BibitemShut {NoStop}%
\bibitem [{\citenamefont {Komineas}\ and\ \citenamefont {Papanicolaou}(2015)}]{Komineas2015}%
  \BibitemOpen
  \bibfield  {author} {\bibinfo {author} {\bibfnamefont {S.}~\bibnamefont {Komineas}}\ and\ \bibinfo {author} {\bibfnamefont {N.}~\bibnamefont {Papanicolaou}},\ }\href {https://doi.org/10.1103/PhysRevB.92.174405} {\bibfield  {journal} {\bibinfo  {journal} {Phys. Rev. B}\ }\textbf {\bibinfo {volume} {92}},\ \bibinfo {pages} {174405} (\bibinfo {year} {2015})}\BibitemShut {NoStop}%
\bibitem [{\citenamefont {Woo}\ \emph {et~al.}(2018)\citenamefont {Woo}, \citenamefont {Song}, \citenamefont {Zhang}, \citenamefont {Zhou}, \citenamefont {Ezawa}, \citenamefont {Liu}, \citenamefont {Finizio}, \citenamefont {Raabe}, \citenamefont {Lee}, \citenamefont {Kim}, \citenamefont {Park}, \citenamefont {Kim}, \citenamefont {Kim}, \citenamefont {Lee}, \citenamefont {Lee}, \citenamefont {Choi}, \citenamefont {Min}, \citenamefont {Koo},\ and\ \citenamefont {Chang}}]{Woo2018}%
  \BibitemOpen
  \bibfield  {author} {\bibinfo {author} {\bibfnamefont {S.}~\bibnamefont {Woo}}, \bibinfo {author} {\bibfnamefont {K.~M.}\ \bibnamefont {Song}}, \bibinfo {author} {\bibfnamefont {X.}~\bibnamefont {Zhang}}, \bibinfo {author} {\bibfnamefont {Y.}~\bibnamefont {Zhou}}, \bibinfo {author} {\bibfnamefont {M.}~\bibnamefont {Ezawa}}, \bibinfo {author} {\bibfnamefont {X.}~\bibnamefont {Liu}}, \bibinfo {author} {\bibfnamefont {S.}~\bibnamefont {Finizio}}, \bibinfo {author} {\bibfnamefont {J.}~\bibnamefont {Raabe}}, \bibinfo {author} {\bibfnamefont {N.~J.}\ \bibnamefont {Lee}}, \bibinfo {author} {\bibfnamefont {S.-I.}\ \bibnamefont {Kim}}, \bibinfo {author} {\bibfnamefont {S.-Y.}\ \bibnamefont {Park}}, \bibinfo {author} {\bibfnamefont {Y.}~\bibnamefont {Kim}}, \bibinfo {author} {\bibfnamefont {J.-Y.}\ \bibnamefont {Kim}}, \bibinfo {author} {\bibfnamefont {D.}~\bibnamefont {Lee}}, \bibinfo {author} {\bibfnamefont {O.}~\bibnamefont {Lee}}, \bibinfo {author} {\bibfnamefont {J.~W.}\ \bibnamefont {Choi}}, \bibinfo {author} {\bibfnamefont {B.-C.}\ \bibnamefont {Min}}, \bibinfo {author} {\bibfnamefont {H.~C.}\ \bibnamefont {Koo}},\ and\ \bibinfo {author} {\bibfnamefont {J.}~\bibnamefont {Chang}},\ }\href {https://doi.org/10.1038/s41467-018-03378-7} {\bibfield  {journal} {\bibinfo  {journal} {Nat. Commun.}\ }\textbf {\bibinfo {volume} {9}},\ \bibinfo {pages} {959} (\bibinfo {year} {2018})}\BibitemShut {NoStop}%
\bibitem [{\citenamefont {Zhang}\ \emph {et~al.}(2020)\citenamefont {Zhang}, \citenamefont {Xia}, \citenamefont {Shen}, \citenamefont {Ezawa}, \citenamefont {Tretiakov}, \citenamefont {Zhao}, \citenamefont {Liu},\ and\ \citenamefont {Zhou}}]{Zhang2020a}%
  \BibitemOpen
  \bibfield  {author} {\bibinfo {author} {\bibfnamefont {X.}~\bibnamefont {Zhang}}, \bibinfo {author} {\bibfnamefont {J.}~\bibnamefont {Xia}}, \bibinfo {author} {\bibfnamefont {L.}~\bibnamefont {Shen}}, \bibinfo {author} {\bibfnamefont {M.}~\bibnamefont {Ezawa}}, \bibinfo {author} {\bibfnamefont {O.~A.}\ \bibnamefont {Tretiakov}}, \bibinfo {author} {\bibfnamefont {G.}~\bibnamefont {Zhao}}, \bibinfo {author} {\bibfnamefont {X.}~\bibnamefont {Liu}},\ and\ \bibinfo {author} {\bibfnamefont {Y.}~\bibnamefont {Zhou}},\ }\href {https://doi.org/10.1103/PhysRevB.101.144435} {\bibfield  {journal} {\bibinfo  {journal} {Phys. Rev. B}\ }\textbf {\bibinfo {volume} {101}},\ \bibinfo {pages} {144435} (\bibinfo {year} {2020})}\BibitemShut {NoStop}%
\bibitem [{\citenamefont {Xia}\ \emph {et~al.}(2019)\citenamefont {Xia}, \citenamefont {Zhang}, \citenamefont {Ezawa}, \citenamefont {Hou}, \citenamefont {Wang}, \citenamefont {Liu},\ and\ \citenamefont {Zhou}}]{Xia2019}%
  \BibitemOpen
  \bibfield  {author} {\bibinfo {author} {\bibfnamefont {J.}~\bibnamefont {Xia}}, \bibinfo {author} {\bibfnamefont {X.}~\bibnamefont {Zhang}}, \bibinfo {author} {\bibfnamefont {M.}~\bibnamefont {Ezawa}}, \bibinfo {author} {\bibfnamefont {Z.}~\bibnamefont {Hou}}, \bibinfo {author} {\bibfnamefont {W.}~\bibnamefont {Wang}}, \bibinfo {author} {\bibfnamefont {X.}~\bibnamefont {Liu}},\ and\ \bibinfo {author} {\bibfnamefont {Y.}~\bibnamefont {Zhou}},\ }\href {https://doi.org/10.1103/PhysRevApplied.11.044046} {\bibfield  {journal} {\bibinfo  {journal} {Phys. Rev. Appl.}\ }\textbf {\bibinfo {volume} {11}},\ \bibinfo {pages} {044046} (\bibinfo {year} {2019})}\BibitemShut {NoStop}%
\bibitem [{\citenamefont {Hou}\ \emph {et~al.}(2020)\citenamefont {Hou}, \citenamefont {Zhang}, \citenamefont {Zhang}, \citenamefont {Xu}, \citenamefont {Xia}, \citenamefont {Ding}, \citenamefont {Li}, \citenamefont {Zhang}, \citenamefont {Batra}, \citenamefont {Costa}, \citenamefont {Liu}, \citenamefont {Wu}, \citenamefont {Ezawa}, \citenamefont {Liu}, \citenamefont {Zhou}, \citenamefont {Zhang},\ and\ \citenamefont {Wang}}]{Hou2020}%
  \BibitemOpen
  \bibfield  {author} {\bibinfo {author} {\bibfnamefont {Z.}~\bibnamefont {Hou}}, \bibinfo {author} {\bibfnamefont {Q.}~\bibnamefont {Zhang}}, \bibinfo {author} {\bibfnamefont {X.}~\bibnamefont {Zhang}}, \bibinfo {author} {\bibfnamefont {G.}~\bibnamefont {Xu}}, \bibinfo {author} {\bibfnamefont {J.}~\bibnamefont {Xia}}, \bibinfo {author} {\bibfnamefont {B.}~\bibnamefont {Ding}}, \bibinfo {author} {\bibfnamefont {H.}~\bibnamefont {Li}}, \bibinfo {author} {\bibfnamefont {S.}~\bibnamefont {Zhang}}, \bibinfo {author} {\bibfnamefont {N.~M.}\ \bibnamefont {Batra}}, \bibinfo {author} {\bibfnamefont {P.~M. F.~J.}\ \bibnamefont {Costa}}, \bibinfo {author} {\bibfnamefont {E.}~\bibnamefont {Liu}}, \bibinfo {author} {\bibfnamefont {G.}~\bibnamefont {Wu}}, \bibinfo {author} {\bibfnamefont {M.}~\bibnamefont {Ezawa}}, \bibinfo {author} {\bibfnamefont {X.}~\bibnamefont {Liu}}, \bibinfo {author} {\bibfnamefont {Y.}~\bibnamefont {Zhou}}, \bibinfo {author} {\bibfnamefont {X.}~\bibnamefont {Zhang}},\ and\ \bibinfo {author} {\bibfnamefont {W.}~\bibnamefont {Wang}},\ }\href {https://doi.org/10.1002/adma.201904815} {\bibfield  {journal} {\bibinfo  {journal} {Adv. Mater.}\ }\textbf {\bibinfo {volume} {32}},\ \bibinfo {pages} {1904815} (\bibinfo {year} {2020})}\BibitemShut {NoStop}%
\bibitem [{\citenamefont {Moon}\ \emph {et~al.}(2015)\citenamefont {Moon}, \citenamefont {Kim}, \citenamefont {Yoo}, \citenamefont {Je}, \citenamefont {Chun}, \citenamefont {Kim}, \citenamefont {Min}, \citenamefont {Hwang},\ and\ \citenamefont {Choe}}]{Moon2015}%
  \BibitemOpen
  \bibfield  {author} {\bibinfo {author} {\bibfnamefont {K.-W.}\ \bibnamefont {Moon}}, \bibinfo {author} {\bibfnamefont {D.-H.}\ \bibnamefont {Kim}}, \bibinfo {author} {\bibfnamefont {S.-C.}\ \bibnamefont {Yoo}}, \bibinfo {author} {\bibfnamefont {S.-G.}\ \bibnamefont {Je}}, \bibinfo {author} {\bibfnamefont {B.~S.}\ \bibnamefont {Chun}}, \bibinfo {author} {\bibfnamefont {W.}~\bibnamefont {Kim}}, \bibinfo {author} {\bibfnamefont {B.-C.}\ \bibnamefont {Min}}, \bibinfo {author} {\bibfnamefont {C.}~\bibnamefont {Hwang}},\ and\ \bibinfo {author} {\bibfnamefont {S.-B.}\ \bibnamefont {Choe}},\ }\href {https://doi.org/10.1038/srep09166} {\bibfield  {journal} {\bibinfo  {journal} {Sci. Rep.}\ }\textbf {\bibinfo {volume} {5}},\ \bibinfo {pages} {9166} (\bibinfo {year} {2015})}\BibitemShut {NoStop}%
\bibitem [{\citenamefont {Zhang}\ \emph {et~al.}(2018{\natexlab{a}})\citenamefont {Zhang}, \citenamefont {Wang}, \citenamefont {Burn}, \citenamefont {Peng}, \citenamefont {Berger}, \citenamefont {Bauer}, \citenamefont {Pfleiderer}, \citenamefont {van~der Laan},\ and\ \citenamefont {Hesjedal}}]{Zhang2018}%
  \BibitemOpen
  \bibfield  {author} {\bibinfo {author} {\bibfnamefont {S.~L.}\ \bibnamefont {Zhang}}, \bibinfo {author} {\bibfnamefont {W.~W.}\ \bibnamefont {Wang}}, \bibinfo {author} {\bibfnamefont {D.~M.}\ \bibnamefont {Burn}}, \bibinfo {author} {\bibfnamefont {H.}~\bibnamefont {Peng}}, \bibinfo {author} {\bibfnamefont {H.}~\bibnamefont {Berger}}, \bibinfo {author} {\bibfnamefont {A.}~\bibnamefont {Bauer}}, \bibinfo {author} {\bibfnamefont {C.}~\bibnamefont {Pfleiderer}}, \bibinfo {author} {\bibfnamefont {G.}~\bibnamefont {van~der Laan}},\ and\ \bibinfo {author} {\bibfnamefont {T.}~\bibnamefont {Hesjedal}},\ }\href {https://doi.org/10.1038/s41467-018-04563-4} {\bibfield  {journal} {\bibinfo  {journal} {Nat. Commun.}\ }\textbf {\bibinfo {volume} {9}},\ \bibinfo {pages} {2115} (\bibinfo {year} {2018}{\natexlab{a}})}\BibitemShut {NoStop}%
\bibitem [{\citenamefont {Raimondo}\ \emph {et~al.}(2022)\citenamefont {Raimondo}, \citenamefont {Saugar}, \citenamefont {Barker}, \citenamefont {Rodrigues}, \citenamefont {Giordano}, \citenamefont {Carpentieri}, \citenamefont {Jiang}, \citenamefont {Chubykalo-Fesenko}, \citenamefont {Tomasello},\ and\ \citenamefont {Finocchio}}]{Raimondo2022}%
  \BibitemOpen
  \bibfield  {author} {\bibinfo {author} {\bibfnamefont {E.}~\bibnamefont {Raimondo}}, \bibinfo {author} {\bibfnamefont {E.}~\bibnamefont {Saugar}}, \bibinfo {author} {\bibfnamefont {J.}~\bibnamefont {Barker}}, \bibinfo {author} {\bibfnamefont {D.}~\bibnamefont {Rodrigues}}, \bibinfo {author} {\bibfnamefont {A.}~\bibnamefont {Giordano}}, \bibinfo {author} {\bibfnamefont {M.}~\bibnamefont {Carpentieri}}, \bibinfo {author} {\bibfnamefont {W.}~\bibnamefont {Jiang}}, \bibinfo {author} {\bibfnamefont {O.}~\bibnamefont {Chubykalo-Fesenko}}, \bibinfo {author} {\bibfnamefont {R.}~\bibnamefont {Tomasello}},\ and\ \bibinfo {author} {\bibfnamefont {G.}~\bibnamefont {Finocchio}},\ }\href {https://doi.org/10.1103/PhysRevApplied.18.024062} {\bibfield  {journal} {\bibinfo  {journal} {Phys. Rev. Appl.}\ }\textbf {\bibinfo {volume} {18}},\ \bibinfo {pages} {024062} (\bibinfo {year} {2022})}\BibitemShut {NoStop}%
\bibitem [{\citenamefont {Tengdin}\ \emph {et~al.}(2022)\citenamefont {Tengdin}, \citenamefont {Truc}, \citenamefont {Sapozhnik}, \citenamefont {Kong}, \citenamefont {del Ser}, \citenamefont {Gargiulo}, \citenamefont {Madan}, \citenamefont {Sch\"onenberger}, \citenamefont {Baral}, \citenamefont {Che}, \citenamefont {Magrez}, \citenamefont {Grundler}, \citenamefont {R\o{}nnow}, \citenamefont {Lagrange}, \citenamefont {Zang}, \citenamefont {Rosch},\ and\ \citenamefont {Carbone}}]{Tengdin2022}%
  \BibitemOpen
  \bibfield  {author} {\bibinfo {author} {\bibfnamefont {P.}~\bibnamefont {Tengdin}}, \bibinfo {author} {\bibfnamefont {B.}~\bibnamefont {Truc}}, \bibinfo {author} {\bibfnamefont {A.}~\bibnamefont {Sapozhnik}}, \bibinfo {author} {\bibfnamefont {L.}~\bibnamefont {Kong}}, \bibinfo {author} {\bibfnamefont {N.}~\bibnamefont {del Ser}}, \bibinfo {author} {\bibfnamefont {S.}~\bibnamefont {Gargiulo}}, \bibinfo {author} {\bibfnamefont {I.}~\bibnamefont {Madan}}, \bibinfo {author} {\bibfnamefont {T.}~\bibnamefont {Sch\"onenberger}}, \bibinfo {author} {\bibfnamefont {P.~R.}\ \bibnamefont {Baral}}, \bibinfo {author} {\bibfnamefont {P.}~\bibnamefont {Che}}, \bibinfo {author} {\bibfnamefont {A.}~\bibnamefont {Magrez}}, \bibinfo {author} {\bibfnamefont {D.}~\bibnamefont {Grundler}}, \bibinfo {author} {\bibfnamefont {H.~M.}\ \bibnamefont {R\o{}nnow}}, \bibinfo {author} {\bibfnamefont {T.}~\bibnamefont {Lagrange}}, \bibinfo {author} {\bibfnamefont {J.}~\bibnamefont {Zang}}, \bibinfo {author} {\bibfnamefont {A.}~\bibnamefont {Rosch}},\ and\ \bibinfo {author} {\bibfnamefont {F.}~\bibnamefont {Carbone}},\ }\href {https://doi.org/10.1103/PhysRevX.12.041030} {\bibfield  {journal} {\bibinfo  {journal} {Phys. Rev. X}\ }\textbf {\bibinfo {volume} {12}},\ \bibinfo {pages} {041030} (\bibinfo {year} {2022})}\BibitemShut {NoStop}%
\bibitem [{\citenamefont {Everschor}\ \emph {et~al.}(2012)\citenamefont {Everschor}, \citenamefont {Garst}, \citenamefont {Binz}, \citenamefont {Jonietz}, \citenamefont {M\"uhlbauer}, \citenamefont {Pfleiderer},\ and\ \citenamefont {Rosch}}]{Everschor2012}%
  \BibitemOpen
  \bibfield  {author} {\bibinfo {author} {\bibfnamefont {K.}~\bibnamefont {Everschor}}, \bibinfo {author} {\bibfnamefont {M.}~\bibnamefont {Garst}}, \bibinfo {author} {\bibfnamefont {B.}~\bibnamefont {Binz}}, \bibinfo {author} {\bibfnamefont {F.}~\bibnamefont {Jonietz}}, \bibinfo {author} {\bibfnamefont {S.}~\bibnamefont {M\"uhlbauer}}, \bibinfo {author} {\bibfnamefont {C.}~\bibnamefont {Pfleiderer}},\ and\ \bibinfo {author} {\bibfnamefont {A.}~\bibnamefont {Rosch}},\ }\href {https://doi.org/10.1103/PhysRevB.86.054432} {\bibfield  {journal} {\bibinfo  {journal} {Phys. Rev. B}\ }\textbf {\bibinfo {volume} {86}},\ \bibinfo {pages} {054432} (\bibinfo {year} {2012})}\BibitemShut {NoStop}%
\bibitem [{\citenamefont {Tang}\ \emph {et~al.}(2021)\citenamefont {Tang}, \citenamefont {Wu}, \citenamefont {Wang}, \citenamefont {Kong}, \citenamefont {Lv}, \citenamefont {Wei}, \citenamefont {Zang}, \citenamefont {Tian},\ and\ \citenamefont {Du}}]{Tang2021}%
  \BibitemOpen
  \bibfield  {author} {\bibinfo {author} {\bibfnamefont {J.}~\bibnamefont {Tang}}, \bibinfo {author} {\bibfnamefont {Y.}~\bibnamefont {Wu}}, \bibinfo {author} {\bibfnamefont {W.}~\bibnamefont {Wang}}, \bibinfo {author} {\bibfnamefont {L.}~\bibnamefont {Kong}}, \bibinfo {author} {\bibfnamefont {B.}~\bibnamefont {Lv}}, \bibinfo {author} {\bibfnamefont {W.}~\bibnamefont {Wei}}, \bibinfo {author} {\bibfnamefont {J.}~\bibnamefont {Zang}}, \bibinfo {author} {\bibfnamefont {M.}~\bibnamefont {Tian}},\ and\ \bibinfo {author} {\bibfnamefont {H.}~\bibnamefont {Du}},\ }\href {https://doi.org/10.1038/s41565-021-00954-9} {\bibfield  {journal} {\bibinfo  {journal} {Nat. Nanotechnol.}\ }\textbf {\bibinfo {volume} {16}},\ \bibinfo {pages} {1086} (\bibinfo {year} {2021})}\BibitemShut {NoStop}%
\bibitem [{\citenamefont {Kind}\ and\ \citenamefont {Foster}(2021)}]{Kind2021}%
  \BibitemOpen
  \bibfield  {author} {\bibinfo {author} {\bibfnamefont {C.}~\bibnamefont {Kind}}\ and\ \bibinfo {author} {\bibfnamefont {D.}~\bibnamefont {Foster}},\ }\href {https://doi.org/10.1103/PhysRevB.103.L100413} {\bibfield  {journal} {\bibinfo  {journal} {Phys. Rev. B}\ }\textbf {\bibinfo {volume} {103}},\ \bibinfo {pages} {L100413} (\bibinfo {year} {2021})}\BibitemShut {NoStop}%
\bibitem [{\citenamefont {Nothhelfer}\ \emph {et~al.}(2019)\citenamefont {Nothhelfer}, \citenamefont {Hals}, \citenamefont {Everschor-Sitte},\ and\ \citenamefont {Rizzi}}]{Nothhelferpatent2019}%
  \BibitemOpen
  \bibfield  {author} {\bibinfo {author} {\bibfnamefont {J.}~\bibnamefont {Nothhelfer}}, \bibinfo {author} {\bibfnamefont {K.~M.~D.}\ \bibnamefont {Hals}}, \bibinfo {author} {\bibfnamefont {K.}~\bibnamefont {Everschor-Sitte}},\ and\ \bibinfo {author} {\bibfnamefont {M.}~\bibnamefont {Rizzi}},\ }\href@noop {} {\bibfield  {journal} {\bibinfo  {journal} {European Patent Application EP3751472A1/International Patent Application WO2020EP66120}\ } (\bibinfo {year} {2019})}\BibitemShut {NoStop}%
\bibitem [{\citenamefont {Yu}\ \emph {et~al.}(2012)\citenamefont {Yu}, \citenamefont {Kanazawa}, \citenamefont {Zhang}, \citenamefont {Nagai}, \citenamefont {Hara}, \citenamefont {Kimoto}, \citenamefont {Matsui}, \citenamefont {Onose},\ and\ \citenamefont {Tokura}}]{Yu2012}%
  \BibitemOpen
  \bibfield  {author} {\bibinfo {author} {\bibfnamefont {X.~Z.}\ \bibnamefont {Yu}}, \bibinfo {author} {\bibfnamefont {N.}~\bibnamefont {Kanazawa}}, \bibinfo {author} {\bibfnamefont {W.~Z.}\ \bibnamefont {Zhang}}, \bibinfo {author} {\bibfnamefont {T.}~\bibnamefont {Nagai}}, \bibinfo {author} {\bibfnamefont {T.}~\bibnamefont {Hara}}, \bibinfo {author} {\bibfnamefont {K.}~\bibnamefont {Kimoto}}, \bibinfo {author} {\bibfnamefont {Y.}~\bibnamefont {Matsui}}, \bibinfo {author} {\bibfnamefont {Y.}~\bibnamefont {Onose}},\ and\ \bibinfo {author} {\bibfnamefont {Y.}~\bibnamefont {Tokura}},\ }\href {https://doi.org/10.1038/ncomms1990} {\bibfield  {journal} {\bibinfo  {journal} {Nat. Commun.}\ }\textbf {\bibinfo {volume} {3}},\ \bibinfo {pages} {988} (\bibinfo {year} {2012})}\BibitemShut {NoStop}%
\bibitem [{\citenamefont {Yu}\ \emph {et~al.}(2020)\citenamefont {Yu}, \citenamefont {Morikawa}, \citenamefont {Nakajima}, \citenamefont {Shibata}, \citenamefont {Kanazawa}, \citenamefont {Arima}, \citenamefont {Nagaosa},\ and\ \citenamefont {Tokura}}]{Yu2020}%
  \BibitemOpen
  \bibfield  {author} {\bibinfo {author} {\bibfnamefont {X.~Z.}\ \bibnamefont {Yu}}, \bibinfo {author} {\bibfnamefont {D.}~\bibnamefont {Morikawa}}, \bibinfo {author} {\bibfnamefont {K.}~\bibnamefont {Nakajima}}, \bibinfo {author} {\bibfnamefont {K.}~\bibnamefont {Shibata}}, \bibinfo {author} {\bibfnamefont {N.}~\bibnamefont {Kanazawa}}, \bibinfo {author} {\bibfnamefont {T.}~\bibnamefont {Arima}}, \bibinfo {author} {\bibfnamefont {N.}~\bibnamefont {Nagaosa}},\ and\ \bibinfo {author} {\bibfnamefont {Y.}~\bibnamefont {Tokura}},\ }\href {https://doi.org/10.1126/sciadv.aaz9744} {\bibfield  {journal} {\bibinfo  {journal} {Sci. Adv.}\ }\textbf {\bibinfo {volume} {6}},\ \bibinfo {pages} {eaaz9744} (\bibinfo {year} {2020})}\BibitemShut {NoStop}%
\bibitem [{\citenamefont {Seki}\ \emph {et~al.}(2012{\natexlab{a}})\citenamefont {Seki}, \citenamefont {Kim}, \citenamefont {Inosov}, \citenamefont {Georgii}, \citenamefont {Keimer}, \citenamefont {Ishiwata},\ and\ \citenamefont {Tokura}}]{Seki2012b}%
  \BibitemOpen
  \bibfield  {author} {\bibinfo {author} {\bibfnamefont {S.}~\bibnamefont {Seki}}, \bibinfo {author} {\bibfnamefont {J.-H.}\ \bibnamefont {Kim}}, \bibinfo {author} {\bibfnamefont {D.~S.}\ \bibnamefont {Inosov}}, \bibinfo {author} {\bibfnamefont {R.}~\bibnamefont {Georgii}}, \bibinfo {author} {\bibfnamefont {B.}~\bibnamefont {Keimer}}, \bibinfo {author} {\bibfnamefont {S.}~\bibnamefont {Ishiwata}},\ and\ \bibinfo {author} {\bibfnamefont {Y.}~\bibnamefont {Tokura}},\ }\href {https://doi.org/10.1103/PhysRevB.85.220406} {\bibfield  {journal} {\bibinfo  {journal} {Phys. Rev. B}\ }\textbf {\bibinfo {volume} {85}},\ \bibinfo {pages} {220406} (\bibinfo {year} {2012}{\natexlab{a}})}\BibitemShut {NoStop}%
\bibitem [{\citenamefont {White}\ \emph {et~al.}(2014)\citenamefont {White}, \citenamefont {Pr\ifmmode~\check{s}\else \v{s}\fi{}a}, \citenamefont {Huang}, \citenamefont {Omrani}, \citenamefont {\ifmmode \check{Z}\else \v{Z}\fi{}ivkovi\ifmmode~\acute{c}\else \'{c}\fi{}}, \citenamefont {Bartkowiak}, \citenamefont {Berger}, \citenamefont {Magrez}, \citenamefont {Gavilano}, \citenamefont {Nagy}, \citenamefont {Zang},\ and\ \citenamefont {R\o{}nnow}}]{White2014}%
  \BibitemOpen
  \bibfield  {author} {\bibinfo {author} {\bibfnamefont {J.~S.}\ \bibnamefont {White}}, \bibinfo {author} {\bibfnamefont {K.}~\bibnamefont {Pr\ifmmode~\check{s}\else \v{s}\fi{}a}}, \bibinfo {author} {\bibfnamefont {P.}~\bibnamefont {Huang}}, \bibinfo {author} {\bibfnamefont {A.~A.}\ \bibnamefont {Omrani}}, \bibinfo {author} {\bibfnamefont {I.}~\bibnamefont {\ifmmode \check{Z}\else \v{Z}\fi{}ivkovi\ifmmode~\acute{c}\else \'{c}\fi{}}}, \bibinfo {author} {\bibfnamefont {M.}~\bibnamefont {Bartkowiak}}, \bibinfo {author} {\bibfnamefont {H.}~\bibnamefont {Berger}}, \bibinfo {author} {\bibfnamefont {A.}~\bibnamefont {Magrez}}, \bibinfo {author} {\bibfnamefont {J.~L.}\ \bibnamefont {Gavilano}}, \bibinfo {author} {\bibfnamefont {G.}~\bibnamefont {Nagy}}, \bibinfo {author} {\bibfnamefont {J.}~\bibnamefont {Zang}},\ and\ \bibinfo {author} {\bibfnamefont {H.~M.}\ \bibnamefont {R\o{}nnow}},\ }\href {https://doi.org/10.1103/PhysRevLett.113.107203} {\bibfield  {journal} {\bibinfo  {journal} {Phys. Rev. Lett.}\ }\textbf {\bibinfo {volume} {113}},\ \bibinfo {pages} {107203} (\bibinfo {year} {2014})}\BibitemShut {NoStop}%
\bibitem [{\citenamefont {Zhang}\ \emph {et~al.}(2018{\natexlab{b}})\citenamefont {Zhang}, \citenamefont {Wang}, \citenamefont {Burn}, \citenamefont {Peng}, \citenamefont {Berger}, \citenamefont {Bauer}, \citenamefont {Pfleiderer}, \citenamefont {{van der Laan}},\ and\ \citenamefont {Hesjedal}}]{Zhang2018c}%
  \BibitemOpen
  \bibfield  {author} {\bibinfo {author} {\bibfnamefont {S.~L.}\ \bibnamefont {Zhang}}, \bibinfo {author} {\bibfnamefont {W.~W.}\ \bibnamefont {Wang}}, \bibinfo {author} {\bibfnamefont {D.~M.}\ \bibnamefont {Burn}}, \bibinfo {author} {\bibfnamefont {H.}~\bibnamefont {Peng}}, \bibinfo {author} {\bibfnamefont {H.}~\bibnamefont {Berger}}, \bibinfo {author} {\bibfnamefont {A.}~\bibnamefont {Bauer}}, \bibinfo {author} {\bibfnamefont {C.}~\bibnamefont {Pfleiderer}}, \bibinfo {author} {\bibfnamefont {G.}~\bibnamefont {{van der Laan}}},\ and\ \bibinfo {author} {\bibfnamefont {T.}~\bibnamefont {Hesjedal}},\ }\href {https://doi.org/10.1038/s41467-018-04563-4} {\bibfield  {journal} {\bibinfo  {journal} {Nat. Commun.}\ }\textbf {\bibinfo {volume} {9}},\ \bibinfo {pages} {2115} (\bibinfo {year} {2018}{\natexlab{b}})}\BibitemShut {NoStop}%
\bibitem [{\citenamefont {Ding}\ \emph {et~al.}(2020)\citenamefont {Ding}, \citenamefont {Li}, \citenamefont {Xu}, \citenamefont {Li}, \citenamefont {Hou}, \citenamefont {Liu}, \citenamefont {Xi}, \citenamefont {Xu}, \citenamefont {Yao},\ and\ \citenamefont {Wang}}]{Ding2020}%
  \BibitemOpen
  \bibfield  {author} {\bibinfo {author} {\bibfnamefont {B.}~\bibnamefont {Ding}}, \bibinfo {author} {\bibfnamefont {Z.}~\bibnamefont {Li}}, \bibinfo {author} {\bibfnamefont {G.}~\bibnamefont {Xu}}, \bibinfo {author} {\bibfnamefont {H.}~\bibnamefont {Li}}, \bibinfo {author} {\bibfnamefont {Z.}~\bibnamefont {Hou}}, \bibinfo {author} {\bibfnamefont {E.}~\bibnamefont {Liu}}, \bibinfo {author} {\bibfnamefont {X.}~\bibnamefont {Xi}}, \bibinfo {author} {\bibfnamefont {F.}~\bibnamefont {Xu}}, \bibinfo {author} {\bibfnamefont {Y.}~\bibnamefont {Yao}},\ and\ \bibinfo {author} {\bibfnamefont {W.}~\bibnamefont {Wang}},\ }\href {https://doi.org/10.1021/acs.nanolett.9b03453} {\bibfield  {journal} {\bibinfo  {journal} {Nano Lett.}\ }\textbf {\bibinfo {volume} {20}},\ \bibinfo {pages} {868} (\bibinfo {year} {2020})}\BibitemShut {NoStop}%
\bibitem [{\citenamefont {Jiang}\ \emph {et~al.}(2016)\citenamefont {Jiang}, \citenamefont {Zhang}, \citenamefont {Yu}, \citenamefont {Jungfleisch}, \citenamefont {Upadhyaya}, \citenamefont {Somaily}, \citenamefont {Pearson}, \citenamefont {Tserkovnyak}, \citenamefont {Wang}, \citenamefont {Heinonen}, \citenamefont {{te Velthuis}},\ and\ \citenamefont {Hoffmann}}]{Jiang2016}%
  \BibitemOpen
  \bibfield  {author} {\bibinfo {author} {\bibfnamefont {W.}~\bibnamefont {Jiang}}, \bibinfo {author} {\bibfnamefont {W.}~\bibnamefont {Zhang}}, \bibinfo {author} {\bibfnamefont {G.}~\bibnamefont {Yu}}, \bibinfo {author} {\bibfnamefont {M.~B.}\ \bibnamefont {Jungfleisch}}, \bibinfo {author} {\bibfnamefont {P.}~\bibnamefont {Upadhyaya}}, \bibinfo {author} {\bibfnamefont {H.}~\bibnamefont {Somaily}}, \bibinfo {author} {\bibfnamefont {J.~E.}\ \bibnamefont {Pearson}}, \bibinfo {author} {\bibfnamefont {Y.}~\bibnamefont {Tserkovnyak}}, \bibinfo {author} {\bibfnamefont {K.~L.}\ \bibnamefont {Wang}}, \bibinfo {author} {\bibfnamefont {O.}~\bibnamefont {Heinonen}}, \bibinfo {author} {\bibfnamefont {S.~G.~E.}\ \bibnamefont {{te Velthuis}}},\ and\ \bibinfo {author} {\bibfnamefont {A.}~\bibnamefont {Hoffmann}},\ }\href {https://doi.org/10.1063/1.4943757} {\bibfield  {journal} {\bibinfo  {journal} {AIP Adv.}\ }\textbf {\bibinfo {volume} {6}},\ \bibinfo {pages} {055602} (\bibinfo {year} {2016})}\BibitemShut {NoStop}%
\bibitem [{\citenamefont {Woo}\ \emph {et~al.}(2016)\citenamefont {Woo}, \citenamefont {Litzius}, \citenamefont {Kr{\"u}ger}, \citenamefont {Im}, \citenamefont {Caretta}, \citenamefont {Richter}, \citenamefont {Mann}, \citenamefont {Krone}, \citenamefont {Reeve}, \citenamefont {Weigand}, \citenamefont {Agrawal}, \citenamefont {Lemesh}, \citenamefont {Mawass}, \citenamefont {Fischer}, \citenamefont {Kl{\"a}ui},\ and\ \citenamefont {Beach}}]{Woo2016}%
  \BibitemOpen
  \bibfield  {author} {\bibinfo {author} {\bibfnamefont {S.}~\bibnamefont {Woo}}, \bibinfo {author} {\bibfnamefont {K.}~\bibnamefont {Litzius}}, \bibinfo {author} {\bibfnamefont {B.}~\bibnamefont {Kr{\"u}ger}}, \bibinfo {author} {\bibfnamefont {M.-Y.}\ \bibnamefont {Im}}, \bibinfo {author} {\bibfnamefont {L.}~\bibnamefont {Caretta}}, \bibinfo {author} {\bibfnamefont {K.}~\bibnamefont {Richter}}, \bibinfo {author} {\bibfnamefont {M.}~\bibnamefont {Mann}}, \bibinfo {author} {\bibfnamefont {A.}~\bibnamefont {Krone}}, \bibinfo {author} {\bibfnamefont {R.~M.}\ \bibnamefont {Reeve}}, \bibinfo {author} {\bibfnamefont {M.}~\bibnamefont {Weigand}}, \bibinfo {author} {\bibfnamefont {P.}~\bibnamefont {Agrawal}}, \bibinfo {author} {\bibfnamefont {I.}~\bibnamefont {Lemesh}}, \bibinfo {author} {\bibfnamefont {M.-A.}\ \bibnamefont {Mawass}}, \bibinfo {author} {\bibfnamefont {P.}~\bibnamefont {Fischer}}, \bibinfo {author} {\bibfnamefont {M.}~\bibnamefont {Kl{\"a}ui}},\ and\ \bibinfo {author} {\bibfnamefont {G.~S.~D.}\ \bibnamefont {Beach}},\ }\href {https://doi.org/10.1038/nmat4593} {\bibfield  {journal} {\bibinfo  {journal} {Nat. Mater.}\ }\textbf {\bibinfo {volume} {15}},\ \bibinfo {pages} {501} (\bibinfo {year} {2016})}\BibitemShut {NoStop}%
\bibitem [{\citenamefont {Kurumaji}\ \emph {et~al.}(2019)\citenamefont {Kurumaji}, \citenamefont {Nakajima}, \citenamefont {Hirschberger}, \citenamefont {Kikkawa}, \citenamefont {Yamasaki}, \citenamefont {Sagayama}, \citenamefont {Nakao}, \citenamefont {Taguchi}, \citenamefont {Arima},\ and\ \citenamefont {Tokura}}]{Kurumaji2019}%
  \BibitemOpen
  \bibfield  {author} {\bibinfo {author} {\bibfnamefont {T.}~\bibnamefont {Kurumaji}}, \bibinfo {author} {\bibfnamefont {T.}~\bibnamefont {Nakajima}}, \bibinfo {author} {\bibfnamefont {M.}~\bibnamefont {Hirschberger}}, \bibinfo {author} {\bibfnamefont {A.}~\bibnamefont {Kikkawa}}, \bibinfo {author} {\bibfnamefont {Y.}~\bibnamefont {Yamasaki}}, \bibinfo {author} {\bibfnamefont {H.}~\bibnamefont {Sagayama}}, \bibinfo {author} {\bibfnamefont {H.}~\bibnamefont {Nakao}}, \bibinfo {author} {\bibfnamefont {Y.}~\bibnamefont {Taguchi}}, \bibinfo {author} {\bibfnamefont {T.-h.}\ \bibnamefont {Arima}},\ and\ \bibinfo {author} {\bibfnamefont {Y.}~\bibnamefont {Tokura}},\ }\href {https://doi.org/10.1126/science.aau0968} {\bibfield  {journal} {\bibinfo  {journal} {Science}\ }\textbf {\bibinfo {volume} {365}},\ \bibinfo {pages} {914} (\bibinfo {year} {2019})}\BibitemShut {NoStop}%
\bibitem [{\citenamefont {Khanh}\ \emph {et~al.}(2020)\citenamefont {Khanh}, \citenamefont {Nakajima}, \citenamefont {Yu}, \citenamefont {Gao}, \citenamefont {Shibata}, \citenamefont {Hirschberger}, \citenamefont {Yamasaki}, \citenamefont {Sagayama}, \citenamefont {Nakao}, \citenamefont {Peng}, \citenamefont {Nakajima}, \citenamefont {Takagi}, \citenamefont {Arima}, \citenamefont {Tokura},\ and\ \citenamefont {Seki}}]{Khanh2020}%
  \BibitemOpen
  \bibfield  {author} {\bibinfo {author} {\bibfnamefont {N.~D.}\ \bibnamefont {Khanh}}, \bibinfo {author} {\bibfnamefont {T.}~\bibnamefont {Nakajima}}, \bibinfo {author} {\bibfnamefont {X.}~\bibnamefont {Yu}}, \bibinfo {author} {\bibfnamefont {S.}~\bibnamefont {Gao}}, \bibinfo {author} {\bibfnamefont {K.}~\bibnamefont {Shibata}}, \bibinfo {author} {\bibfnamefont {M.}~\bibnamefont {Hirschberger}}, \bibinfo {author} {\bibfnamefont {Y.}~\bibnamefont {Yamasaki}}, \bibinfo {author} {\bibfnamefont {H.}~\bibnamefont {Sagayama}}, \bibinfo {author} {\bibfnamefont {H.}~\bibnamefont {Nakao}}, \bibinfo {author} {\bibfnamefont {L.}~\bibnamefont {Peng}}, \bibinfo {author} {\bibfnamefont {K.}~\bibnamefont {Nakajima}}, \bibinfo {author} {\bibfnamefont {R.}~\bibnamefont {Takagi}}, \bibinfo {author} {\bibfnamefont {T.-h.}\ \bibnamefont {Arima}}, \bibinfo {author} {\bibfnamefont {Y.}~\bibnamefont {Tokura}},\ and\ \bibinfo {author} {\bibfnamefont {S.}~\bibnamefont {Seki}},\ }\href {https://doi.org/10.1038/s41565-020-0684-7} {\bibfield  {journal} {\bibinfo  {journal} {Nat. Nanotechnol.}\ }\textbf {\bibinfo {volume} {15}},\ \bibinfo {pages} {444} (\bibinfo {year} {2020})}\BibitemShut {NoStop}%
\bibitem [{\citenamefont {McKeever}\ \emph {et~al.}(2019)\citenamefont {McKeever}, \citenamefont {Rodrigues}, \citenamefont {Pinna}, \citenamefont {Abanov}, \citenamefont {Sinova},\ and\ \citenamefont {Everschor-Sitte}}]{McKeever2019}%
  \BibitemOpen
  \bibfield  {author} {\bibinfo {author} {\bibfnamefont {B.~F.}\ \bibnamefont {McKeever}}, \bibinfo {author} {\bibfnamefont {D.~R.}\ \bibnamefont {Rodrigues}}, \bibinfo {author} {\bibfnamefont {D.}~\bibnamefont {Pinna}}, \bibinfo {author} {\bibfnamefont {A.}~\bibnamefont {Abanov}}, \bibinfo {author} {\bibfnamefont {J.}~\bibnamefont {Sinova}},\ and\ \bibinfo {author} {\bibfnamefont {K.}~\bibnamefont {Everschor-Sitte}},\ }\href {https://doi.org/10.1103/PhysRevB.99.054430} {\bibfield  {journal} {\bibinfo  {journal} {Phys. Rev. B}\ }\textbf {\bibinfo {volume} {99}},\ \bibinfo {pages} {054430} (\bibinfo {year} {2019})}\BibitemShut {NoStop}%
\bibitem [{\citenamefont {Zhou}\ \emph {et~al.}(2015)\citenamefont {Zhou}, \citenamefont {Iacocca}, \citenamefont {Awad}, \citenamefont {Dumas}, \citenamefont {Zhang}, \citenamefont {Braun},\ and\ \citenamefont {{\AA}kerman}}]{Zhou2015}%
  \BibitemOpen
  \bibfield  {author} {\bibinfo {author} {\bibfnamefont {Y.}~\bibnamefont {Zhou}}, \bibinfo {author} {\bibfnamefont {E.}~\bibnamefont {Iacocca}}, \bibinfo {author} {\bibfnamefont {A.~A.}\ \bibnamefont {Awad}}, \bibinfo {author} {\bibfnamefont {R.~K.}\ \bibnamefont {Dumas}}, \bibinfo {author} {\bibfnamefont {F.~C.}\ \bibnamefont {Zhang}}, \bibinfo {author} {\bibfnamefont {H.~B.}\ \bibnamefont {Braun}},\ and\ \bibinfo {author} {\bibfnamefont {J.}~\bibnamefont {{\AA}kerman}},\ }\href {https://doi.org/10.1038/ncomms9193} {\bibfield  {journal} {\bibinfo  {journal} {Nat. Commun.}\ }\textbf {\bibinfo {volume} {6}},\ \bibinfo {pages} {8193} (\bibinfo {year} {2015})}\BibitemShut {NoStop}%
\bibitem [{\citenamefont {Winter}(1961)}]{Winter1961}%
  \BibitemOpen
  \bibfield  {author} {\bibinfo {author} {\bibfnamefont {J.~M.}\ \bibnamefont {Winter}},\ }\href {https://doi.org/10.1103/PhysRev.124.452} {\bibfield  {journal} {\bibinfo  {journal} {Phys. Rev.}\ }\textbf {\bibinfo {volume} {124}},\ \bibinfo {pages} {452} (\bibinfo {year} {1961})}\BibitemShut {NoStop}%
\bibitem [{\citenamefont {Rohart}\ and\ \citenamefont {Thiaville}(2013)}]{Rohart2013}%
  \BibitemOpen
  \bibfield  {author} {\bibinfo {author} {\bibfnamefont {S.}~\bibnamefont {Rohart}}\ and\ \bibinfo {author} {\bibfnamefont {A.}~\bibnamefont {Thiaville}},\ }\href {https://doi.org/10.1103/PhysRevB.88.184422} {\bibfield  {journal} {\bibinfo  {journal} {Phys. Rev. B}\ }\textbf {\bibinfo {volume} {88}},\ \bibinfo {pages} {184422} (\bibinfo {year} {2013})}\BibitemShut {NoStop}%
\bibitem [{\citenamefont {Mostovoy}(2006)}]{Mostovoy2006}%
  \BibitemOpen
  \bibfield  {author} {\bibinfo {author} {\bibfnamefont {M.}~\bibnamefont {Mostovoy}},\ }\href {https://doi.org/10.1103/PhysRevLett.96.067601} {\bibfield  {journal} {\bibinfo  {journal} {Phys. Rev. Lett.}\ }\textbf {\bibinfo {volume} {96}},\ \bibinfo {pages} {067601} (\bibinfo {year} {2006})}\BibitemShut {NoStop}%
\bibitem [{\citenamefont {Cheong}\ and\ \citenamefont {Mostovoy}(2007)}]{Cheong2007}%
  \BibitemOpen
  \bibfield  {author} {\bibinfo {author} {\bibfnamefont {S.-W.}\ \bibnamefont {Cheong}}\ and\ \bibinfo {author} {\bibfnamefont {M.}~\bibnamefont {Mostovoy}},\ }\href {https://doi.org/10.1038/nmat1804} {\bibfield  {journal} {\bibinfo  {journal} {Nat. Mater.}\ }\textbf {\bibinfo {volume} {6}},\ \bibinfo {pages} {13} (\bibinfo {year} {2007})}\BibitemShut {NoStop}%
\bibitem [{\citenamefont {Seki}\ \emph {et~al.}(2012{\natexlab{b}})\citenamefont {Seki}, \citenamefont {Yu}, \citenamefont {Ishiwata},\ and\ \citenamefont {Tokura}}]{Seki2012}%
  \BibitemOpen
  \bibfield  {author} {\bibinfo {author} {\bibfnamefont {S.}~\bibnamefont {Seki}}, \bibinfo {author} {\bibfnamefont {X.~Z.}\ \bibnamefont {Yu}}, \bibinfo {author} {\bibfnamefont {S.}~\bibnamefont {Ishiwata}},\ and\ \bibinfo {author} {\bibfnamefont {Y.}~\bibnamefont {Tokura}},\ }\href {https://doi.org/10.1126/science.1214143} {\bibfield  {journal} {\bibinfo  {journal} {Science}\ }\textbf {\bibinfo {volume} {336}},\ \bibinfo {pages} {198} (\bibinfo {year} {2012}{\natexlab{b}})}\BibitemShut {NoStop}%
\bibitem [{\citenamefont {Yu}\ \emph {et~al.}(2010)\citenamefont {Yu}, \citenamefont {Onose}, \citenamefont {Kanazawa}, \citenamefont {Park}, \citenamefont {Han}, \citenamefont {Matsui}, \citenamefont {Nagaosa},\ and\ \citenamefont {Tokura}}]{Yu2010}%
  \BibitemOpen
  \bibfield  {author} {\bibinfo {author} {\bibfnamefont {X.~Z.}\ \bibnamefont {Yu}}, \bibinfo {author} {\bibfnamefont {Y.}~\bibnamefont {Onose}}, \bibinfo {author} {\bibfnamefont {N.}~\bibnamefont {Kanazawa}}, \bibinfo {author} {\bibfnamefont {J.~H.}\ \bibnamefont {Park}}, \bibinfo {author} {\bibfnamefont {J.~H.}\ \bibnamefont {Han}}, \bibinfo {author} {\bibfnamefont {Y.}~\bibnamefont {Matsui}}, \bibinfo {author} {\bibfnamefont {N.}~\bibnamefont {Nagaosa}},\ and\ \bibinfo {author} {\bibfnamefont {Y.}~\bibnamefont {Tokura}},\ }\href {https://doi.org/10.1038/nature09124} {\bibfield  {journal} {\bibinfo  {journal} {Nature}\ }\textbf {\bibinfo {volume} {465}},\ \bibinfo {pages} {901} (\bibinfo {year} {2010})}\BibitemShut {NoStop}%
\bibitem [{\citenamefont {Park}\ \emph {et~al.}(2014)\citenamefont {Park}, \citenamefont {Yu}, \citenamefont {Aizawa}, \citenamefont {Tanigaki}, \citenamefont {Akashi}, \citenamefont {Takahashi}, \citenamefont {Matsuda}, \citenamefont {Kanazawa}, \citenamefont {Onose}, \citenamefont {Shindo}, \citenamefont {Tonomura},\ and\ \citenamefont {Tokura}}]{Park2014}%
  \BibitemOpen
  \bibfield  {author} {\bibinfo {author} {\bibfnamefont {H.~S.}\ \bibnamefont {Park}}, \bibinfo {author} {\bibfnamefont {X.}~\bibnamefont {Yu}}, \bibinfo {author} {\bibfnamefont {S.}~\bibnamefont {Aizawa}}, \bibinfo {author} {\bibfnamefont {T.}~\bibnamefont {Tanigaki}}, \bibinfo {author} {\bibfnamefont {T.}~\bibnamefont {Akashi}}, \bibinfo {author} {\bibfnamefont {Y.}~\bibnamefont {Takahashi}}, \bibinfo {author} {\bibfnamefont {T.}~\bibnamefont {Matsuda}}, \bibinfo {author} {\bibfnamefont {N.}~\bibnamefont {Kanazawa}}, \bibinfo {author} {\bibfnamefont {Y.}~\bibnamefont {Onose}}, \bibinfo {author} {\bibfnamefont {D.}~\bibnamefont {Shindo}}, \bibinfo {author} {\bibfnamefont {A.}~\bibnamefont {Tonomura}},\ and\ \bibinfo {author} {\bibfnamefont {Y.}~\bibnamefont {Tokura}},\ }\href {https://doi.org/10.1038/nnano.2014.52} {\bibfield  {journal} {\bibinfo  {journal} {Nat. Nanotechnol.}\ }\textbf {\bibinfo {volume} {9}},\ \bibinfo {pages} {337} (\bibinfo {year} {2014})}\BibitemShut {NoStop}%
\bibitem [{\citenamefont {Shibata}\ \emph {et~al.}(2013)\citenamefont {Shibata}, \citenamefont {Yu}, \citenamefont {Hara}, \citenamefont {Morikawa}, \citenamefont {Kanazawa}, \citenamefont {Kimoto}, \citenamefont {Ishiwata}, \citenamefont {Matsui},\ and\ \citenamefont {Tokura}}]{Shibata2013}%
  \BibitemOpen
  \bibfield  {author} {\bibinfo {author} {\bibfnamefont {K.}~\bibnamefont {Shibata}}, \bibinfo {author} {\bibfnamefont {X.~Z.}\ \bibnamefont {Yu}}, \bibinfo {author} {\bibfnamefont {T.}~\bibnamefont {Hara}}, \bibinfo {author} {\bibfnamefont {D.}~\bibnamefont {Morikawa}}, \bibinfo {author} {\bibfnamefont {N.}~\bibnamefont {Kanazawa}}, \bibinfo {author} {\bibfnamefont {K.}~\bibnamefont {Kimoto}}, \bibinfo {author} {\bibfnamefont {S.}~\bibnamefont {Ishiwata}}, \bibinfo {author} {\bibfnamefont {Y.}~\bibnamefont {Matsui}},\ and\ \bibinfo {author} {\bibfnamefont {Y.}~\bibnamefont {Tokura}},\ }\href {https://doi.org/10.1038/nnano.2013.174} {\bibfield  {journal} {\bibinfo  {journal} {Nat. Nanotechnol.}\ }\textbf {\bibinfo {volume} {8}},\ \bibinfo {pages} {723} (\bibinfo {year} {2013})}\BibitemShut {NoStop}%
\bibitem [{\citenamefont {Zhang}\ \emph {et~al.}(2018{\natexlab{c}})\citenamefont {Zhang}, \citenamefont {van~der Laan}, \citenamefont {Wang}, \citenamefont {Haghighirad},\ and\ \citenamefont {Hesjedal}}]{Zhang2018b}%
  \BibitemOpen
  \bibfield  {author} {\bibinfo {author} {\bibfnamefont {S.~L.}\ \bibnamefont {Zhang}}, \bibinfo {author} {\bibfnamefont {G.}~\bibnamefont {van~der Laan}}, \bibinfo {author} {\bibfnamefont {W.~W.}\ \bibnamefont {Wang}}, \bibinfo {author} {\bibfnamefont {A.~A.}\ \bibnamefont {Haghighirad}},\ and\ \bibinfo {author} {\bibfnamefont {T.}~\bibnamefont {Hesjedal}},\ }\href {https://doi.org/10.1103/PhysRevLett.120.227202} {\bibfield  {journal} {\bibinfo  {journal} {Phys. Rev. Lett.}\ }\textbf {\bibinfo {volume} {120}},\ \bibinfo {pages} {227202} (\bibinfo {year} {2018}{\natexlab{c}})}\BibitemShut {NoStop}%
\bibitem [{\citenamefont {Woo}\ \emph {et~al.}(2017)\citenamefont {Woo}, \citenamefont {Song}, \citenamefont {Han}, \citenamefont {Jung}, \citenamefont {Im}, \citenamefont {Lee}, \citenamefont {Song}, \citenamefont {Fischer}, \citenamefont {Hong}, \citenamefont {Choi}, \citenamefont {Min}, \citenamefont {Koo},\ and\ \citenamefont {Chang}}]{Woo2017}%
  \BibitemOpen
  \bibfield  {author} {\bibinfo {author} {\bibfnamefont {S.}~\bibnamefont {Woo}}, \bibinfo {author} {\bibfnamefont {K.~M.}\ \bibnamefont {Song}}, \bibinfo {author} {\bibfnamefont {H.-S.}\ \bibnamefont {Han}}, \bibinfo {author} {\bibfnamefont {M.-S.}\ \bibnamefont {Jung}}, \bibinfo {author} {\bibfnamefont {M.-Y.}\ \bibnamefont {Im}}, \bibinfo {author} {\bibfnamefont {K.-S.}\ \bibnamefont {Lee}}, \bibinfo {author} {\bibfnamefont {K.~S.}\ \bibnamefont {Song}}, \bibinfo {author} {\bibfnamefont {P.}~\bibnamefont {Fischer}}, \bibinfo {author} {\bibfnamefont {J.-I.}\ \bibnamefont {Hong}}, \bibinfo {author} {\bibfnamefont {J.~W.}\ \bibnamefont {Choi}}, \bibinfo {author} {\bibfnamefont {B.-C.}\ \bibnamefont {Min}}, \bibinfo {author} {\bibfnamefont {H.~C.}\ \bibnamefont {Koo}},\ and\ \bibinfo {author} {\bibfnamefont {J.}~\bibnamefont {Chang}},\ }\href {https://doi.org/10.1038/ncomms15573} {\bibfield  {journal} {\bibinfo  {journal} {Nat. Commun.}\ }\textbf {\bibinfo {volume} {8}},\ \bibinfo {pages} {15573} (\bibinfo {year} {2017})}\BibitemShut {NoStop}%
\bibitem [{\citenamefont {Litzius}\ \emph {et~al.}(2017)\citenamefont {Litzius}, \citenamefont {Lemesh}, \citenamefont {Kr{\"u}ger}, \citenamefont {Bassirian}, \citenamefont {Caretta}, \citenamefont {Richter}, \citenamefont {B{\"u}ttner}, \citenamefont {Sato}, \citenamefont {Tretiakov}, \citenamefont {F{\"o}rster}, \citenamefont {Reeve}, \citenamefont {Weigand}, \citenamefont {Bykova}, \citenamefont {Stoll}, \citenamefont {Sch{\"u}tz}, \citenamefont {Beach},\ and\ \citenamefont {Kl{\"a}ui}}]{Litzius2017}%
  \BibitemOpen
  \bibfield  {author} {\bibinfo {author} {\bibfnamefont {K.}~\bibnamefont {Litzius}}, \bibinfo {author} {\bibfnamefont {I.}~\bibnamefont {Lemesh}}, \bibinfo {author} {\bibfnamefont {B.}~\bibnamefont {Kr{\"u}ger}}, \bibinfo {author} {\bibfnamefont {P.}~\bibnamefont {Bassirian}}, \bibinfo {author} {\bibfnamefont {L.}~\bibnamefont {Caretta}}, \bibinfo {author} {\bibfnamefont {K.}~\bibnamefont {Richter}}, \bibinfo {author} {\bibfnamefont {F.}~\bibnamefont {B{\"u}ttner}}, \bibinfo {author} {\bibfnamefont {K.}~\bibnamefont {Sato}}, \bibinfo {author} {\bibfnamefont {O.~A.}\ \bibnamefont {Tretiakov}}, \bibinfo {author} {\bibfnamefont {J.}~\bibnamefont {F{\"o}rster}}, \bibinfo {author} {\bibfnamefont {R.~M.}\ \bibnamefont {Reeve}}, \bibinfo {author} {\bibfnamefont {M.}~\bibnamefont {Weigand}}, \bibinfo {author} {\bibfnamefont {I.}~\bibnamefont {Bykova}}, \bibinfo {author} {\bibfnamefont {H.}~\bibnamefont {Stoll}}, \bibinfo {author} {\bibfnamefont {G.}~\bibnamefont {Sch{\"u}tz}}, \bibinfo {author} {\bibfnamefont {G.~S.~D.}\ \bibnamefont {Beach}},\ and\ \bibinfo {author} {\bibfnamefont {M.}~\bibnamefont {Kl{\"a}ui}},\ }\href {https://doi.org/10.1038/nphys4000} {\bibfield  {journal} {\bibinfo  {journal} {Nat. Phys.}\ }\textbf {\bibinfo {volume} {13}},\ \bibinfo {pages} {170} (\bibinfo {year} {2017})}\BibitemShut {NoStop}%
\bibitem [{\citenamefont {Jiang}\ \emph {et~al.}(2017)\citenamefont {Jiang}, \citenamefont {Zhang}, \citenamefont {Yu}, \citenamefont {Zhang}, \citenamefont {Wang}, \citenamefont {Benjamin~Jungfleisch}, \citenamefont {Pearson}, \citenamefont {Cheng}, \citenamefont {Heinonen}, \citenamefont {Wang}, \citenamefont {Zhou}, \citenamefont {Hoffmann},\ and\ \citenamefont {te~Velthuis}}]{Jiang2017}%
  \BibitemOpen
  \bibfield  {author} {\bibinfo {author} {\bibfnamefont {W.}~\bibnamefont {Jiang}}, \bibinfo {author} {\bibfnamefont {X.}~\bibnamefont {Zhang}}, \bibinfo {author} {\bibfnamefont {G.}~\bibnamefont {Yu}}, \bibinfo {author} {\bibfnamefont {W.}~\bibnamefont {Zhang}}, \bibinfo {author} {\bibfnamefont {X.}~\bibnamefont {Wang}}, \bibinfo {author} {\bibfnamefont {M.}~\bibnamefont {Benjamin~Jungfleisch}}, \bibinfo {author} {\bibfnamefont {J.~E.}\ \bibnamefont {Pearson}}, \bibinfo {author} {\bibfnamefont {X.}~\bibnamefont {Cheng}}, \bibinfo {author} {\bibfnamefont {O.}~\bibnamefont {Heinonen}}, \bibinfo {author} {\bibfnamefont {K.~L.}\ \bibnamefont {Wang}}, \bibinfo {author} {\bibfnamefont {Y.}~\bibnamefont {Zhou}}, \bibinfo {author} {\bibfnamefont {A.}~\bibnamefont {Hoffmann}},\ and\ \bibinfo {author} {\bibfnamefont {S.~G.~E.}\ \bibnamefont {te~Velthuis}},\ }\href {https://doi.org/10.1038/nphys3883} {\bibfield  {journal} {\bibinfo  {journal} {Nat. Phys.}\ }\textbf {\bibinfo {volume} {13}},\ \bibinfo {pages} {162} (\bibinfo {year} {2017})}\BibitemShut {NoStop}%
\bibitem [{\citenamefont {V{\'e}lez}\ \emph {et~al.}(2022)\citenamefont {V{\'e}lez}, \citenamefont {Ruiz-G{\'o}mez}, \citenamefont {Schaab}, \citenamefont {Gradauskaite}, \citenamefont {W{\"o}rnle}, \citenamefont {Welter}, \citenamefont {Jacot}, \citenamefont {Degen}, \citenamefont {Trassin}, \citenamefont {Fiebig},\ and\ \citenamefont {Gambardella}}]{Velez2022}%
  \BibitemOpen
  \bibfield  {author} {\bibinfo {author} {\bibfnamefont {S.}~\bibnamefont {V{\'e}lez}}, \bibinfo {author} {\bibfnamefont {S.}~\bibnamefont {Ruiz-G{\'o}mez}}, \bibinfo {author} {\bibfnamefont {J.}~\bibnamefont {Schaab}}, \bibinfo {author} {\bibfnamefont {E.}~\bibnamefont {Gradauskaite}}, \bibinfo {author} {\bibfnamefont {M.~S.}\ \bibnamefont {W{\"o}rnle}}, \bibinfo {author} {\bibfnamefont {P.}~\bibnamefont {Welter}}, \bibinfo {author} {\bibfnamefont {B.~J.}\ \bibnamefont {Jacot}}, \bibinfo {author} {\bibfnamefont {C.~L.}\ \bibnamefont {Degen}}, \bibinfo {author} {\bibfnamefont {M.}~\bibnamefont {Trassin}}, \bibinfo {author} {\bibfnamefont {M.}~\bibnamefont {Fiebig}},\ and\ \bibinfo {author} {\bibfnamefont {P.}~\bibnamefont {Gambardella}},\ }\href {https://doi.org/10.1038/s41565-022-01144-x} {\bibfield  {journal} {\bibinfo  {journal} {Nat. Nanotechnol.}\ }\textbf {\bibinfo {volume} {17}},\ \bibinfo {pages} {834} (\bibinfo {year} {2022})}\BibitemShut {NoStop}%
\bibitem [{\citenamefont {Quessab}\ \emph {et~al.}(2022)\citenamefont {Quessab}, \citenamefont {Xu}, \citenamefont {Cogulu}, \citenamefont {Finizio}, \citenamefont {Raabe},\ and\ \citenamefont {Kent}}]{Quessab2022}%
  \BibitemOpen
  \bibfield  {author} {\bibinfo {author} {\bibfnamefont {Y.}~\bibnamefont {Quessab}}, \bibinfo {author} {\bibfnamefont {J.-W.}\ \bibnamefont {Xu}}, \bibinfo {author} {\bibfnamefont {E.}~\bibnamefont {Cogulu}}, \bibinfo {author} {\bibfnamefont {S.}~\bibnamefont {Finizio}}, \bibinfo {author} {\bibfnamefont {J.}~\bibnamefont {Raabe}},\ and\ \bibinfo {author} {\bibfnamefont {A.~D.}\ \bibnamefont {Kent}},\ }\href {https://doi.org/10.1021/acs.nanolett.2c01038} {\bibfield  {journal} {\bibinfo  {journal} {Nano Lett.}\ }\textbf {\bibinfo {volume} {22}},\ \bibinfo {pages} {6091} (\bibinfo {year} {2022})}\BibitemShut {NoStop}%
\bibitem [{\citenamefont {Casiraghi}\ \emph {et~al.}(2019)\citenamefont {Casiraghi}, \citenamefont {Corte-Le{\'o}n}, \citenamefont {Vafaee}, \citenamefont {Garcia-Sanchez}, \citenamefont {Durin}, \citenamefont {Pasquale}, \citenamefont {Jakob}, \citenamefont {Kl{\"a}ui},\ and\ \citenamefont {Kazakova}}]{Casiraghi2019}%
  \BibitemOpen
  \bibfield  {author} {\bibinfo {author} {\bibfnamefont {A.}~\bibnamefont {Casiraghi}}, \bibinfo {author} {\bibfnamefont {H.}~\bibnamefont {Corte-Le{\'o}n}}, \bibinfo {author} {\bibfnamefont {M.}~\bibnamefont {Vafaee}}, \bibinfo {author} {\bibfnamefont {F.}~\bibnamefont {Garcia-Sanchez}}, \bibinfo {author} {\bibfnamefont {G.}~\bibnamefont {Durin}}, \bibinfo {author} {\bibfnamefont {M.}~\bibnamefont {Pasquale}}, \bibinfo {author} {\bibfnamefont {G.}~\bibnamefont {Jakob}}, \bibinfo {author} {\bibfnamefont {M.}~\bibnamefont {Kl{\"a}ui}},\ and\ \bibinfo {author} {\bibfnamefont {O.}~\bibnamefont {Kazakova}},\ }\href {https://doi.org/10.1038/s42005-019-0242-5} {\bibfield  {journal} {\bibinfo  {journal} {Commun. Phys.}\ }\textbf {\bibinfo {volume} {2}},\ \bibinfo {pages} {145} (\bibinfo {year} {2019})}\BibitemShut {NoStop}%
\bibitem [{\citenamefont {Meng}\ \emph {et~al.}(2019)\citenamefont {Meng}, \citenamefont {Ahmed}, \citenamefont {Baćani}, \citenamefont {Mandru}, \citenamefont {Zhao}, \citenamefont {Bagués}, \citenamefont {Esser}, \citenamefont {Flores}, \citenamefont {McComb}, \citenamefont {Hug},\ and\ \citenamefont {Yang}}]{Meng2019}%
  \BibitemOpen
  \bibfield  {author} {\bibinfo {author} {\bibfnamefont {K.-Y.}\ \bibnamefont {Meng}}, \bibinfo {author} {\bibfnamefont {A.~S.}\ \bibnamefont {Ahmed}}, \bibinfo {author} {\bibfnamefont {M.}~\bibnamefont {Baćani}}, \bibinfo {author} {\bibfnamefont {A.-O.}\ \bibnamefont {Mandru}}, \bibinfo {author} {\bibfnamefont {X.}~\bibnamefont {Zhao}}, \bibinfo {author} {\bibfnamefont {N.}~\bibnamefont {Bagués}}, \bibinfo {author} {\bibfnamefont {B.~D.}\ \bibnamefont {Esser}}, \bibinfo {author} {\bibfnamefont {J.}~\bibnamefont {Flores}}, \bibinfo {author} {\bibfnamefont {D.~W.}\ \bibnamefont {McComb}}, \bibinfo {author} {\bibfnamefont {H.~J.}\ \bibnamefont {Hug}},\ and\ \bibinfo {author} {\bibfnamefont {F.}~\bibnamefont {Yang}},\ }\href {https://doi.org/10.1021/acs.nanolett.9b00596} {\bibfield  {journal} {\bibinfo  {journal} {Nano Lett.}\ }\textbf {\bibinfo {volume} {19}},\ \bibinfo {pages} {3169} (\bibinfo {year} {2019})}\BibitemShut {NoStop}%
\bibitem [{\citenamefont {Raju}\ \emph {et~al.}(2019)\citenamefont {Raju}, \citenamefont {Yagil}, \citenamefont {Soumyanarayanan}, \citenamefont {Tan}, \citenamefont {Almoalem}, \citenamefont {Ma}, \citenamefont {Auslaender},\ and\ \citenamefont {Panagopoulos}}]{Raju2019}%
  \BibitemOpen
  \bibfield  {author} {\bibinfo {author} {\bibfnamefont {M.}~\bibnamefont {Raju}}, \bibinfo {author} {\bibfnamefont {A.}~\bibnamefont {Yagil}}, \bibinfo {author} {\bibfnamefont {A.}~\bibnamefont {Soumyanarayanan}}, \bibinfo {author} {\bibfnamefont {A.~K.~C.}\ \bibnamefont {Tan}}, \bibinfo {author} {\bibfnamefont {A.}~\bibnamefont {Almoalem}}, \bibinfo {author} {\bibfnamefont {F.}~\bibnamefont {Ma}}, \bibinfo {author} {\bibfnamefont {O.~M.}\ \bibnamefont {Auslaender}},\ and\ \bibinfo {author} {\bibfnamefont {C.}~\bibnamefont {Panagopoulos}},\ }\href {https://doi.org/10.1038/s41467-018-08041-9} {\bibfield  {journal} {\bibinfo  {journal} {Nat. Commun.}\ }\textbf {\bibinfo {volume} {10}},\ \bibinfo {pages} {696} (\bibinfo {year} {2019})}\BibitemShut {NoStop}%
\bibitem [{\citenamefont {Zhang}\ \emph {et~al.}(2022)\citenamefont {Zhang}, \citenamefont {Raftrey}, \citenamefont {Chan}, \citenamefont {Shao}, \citenamefont {Chen}, \citenamefont {Chen}, \citenamefont {Huang}, \citenamefont {Reichanadter}, \citenamefont {Dong}, \citenamefont {Susarla}, \citenamefont {Caretta}, \citenamefont {Chen}, \citenamefont {Yao}, \citenamefont {Fischer}, \citenamefont {Neaton}, \citenamefont {Wu}, \citenamefont {Muller}, \citenamefont {Birgeneau},\ and\ \citenamefont {Ramesh}}]{Zhang2022}%
  \BibitemOpen
  \bibfield  {author} {\bibinfo {author} {\bibfnamefont {H.}~\bibnamefont {Zhang}}, \bibinfo {author} {\bibfnamefont {D.}~\bibnamefont {Raftrey}}, \bibinfo {author} {\bibfnamefont {Y.-T.}\ \bibnamefont {Chan}}, \bibinfo {author} {\bibfnamefont {Y.-T.}\ \bibnamefont {Shao}}, \bibinfo {author} {\bibfnamefont {R.}~\bibnamefont {Chen}}, \bibinfo {author} {\bibfnamefont {X.}~\bibnamefont {Chen}}, \bibinfo {author} {\bibfnamefont {X.}~\bibnamefont {Huang}}, \bibinfo {author} {\bibfnamefont {J.~T.}\ \bibnamefont {Reichanadter}}, \bibinfo {author} {\bibfnamefont {K.}~\bibnamefont {Dong}}, \bibinfo {author} {\bibfnamefont {S.}~\bibnamefont {Susarla}}, \bibinfo {author} {\bibfnamefont {L.}~\bibnamefont {Caretta}}, \bibinfo {author} {\bibfnamefont {Z.}~\bibnamefont {Chen}}, \bibinfo {author} {\bibfnamefont {J.}~\bibnamefont {Yao}}, \bibinfo {author} {\bibfnamefont {P.}~\bibnamefont {Fischer}}, \bibinfo {author} {\bibfnamefont {J.~B.}\ \bibnamefont {Neaton}}, \bibinfo {author} {\bibfnamefont {W.}~\bibnamefont {Wu}}, \bibinfo {author} {\bibfnamefont {D.~A.}\ \bibnamefont {Muller}}, \bibinfo {author} {\bibfnamefont {R.~J.}\ \bibnamefont {Birgeneau}},\ and\ \bibinfo {author} {\bibfnamefont {R.}~\bibnamefont {Ramesh}},\ }\href {https://doi.org/10.1126/sciadv.abm7103} {\bibfield  {journal} {\bibinfo  {journal} {Sci. Adv.}\ }\textbf {\bibinfo {volume} {8}},\ \bibinfo {pages} {eabm7103} (\bibinfo {year} {2022})}\BibitemShut {NoStop}%
\bibitem [{\citenamefont {Landau}\ and\ \citenamefont {Lifshitz}(1992)}]{Landau1935}%
  \BibitemOpen
  \bibfield  {author} {\bibinfo {author} {\bibfnamefont {L.}~\bibnamefont {Landau}}\ and\ \bibinfo {author} {\bibfnamefont {E.}~\bibnamefont {Lifshitz}},\ }in\ \href {https://doi.org/10.1016/B978-0-08-036364-6.50008-9} {\emph {\bibinfo {booktitle} {Perspectives in Theoretical Physics}}},\ \bibinfo {editor} {edited by\ \bibinfo {editor} {\bibfnamefont {L.}~\bibnamefont {Pitaevski}}}\ (\bibinfo  {publisher} {{Pergamon}},\ \bibinfo {address} {{Amsterdam}},\ \bibinfo {year} {1992})\ pp.\ \bibinfo {pages} {51--65}\BibitemShut {NoStop}%
\bibitem [{\citenamefont {Gilbert}(2004)}]{Gilbert2004}%
  \BibitemOpen
  \bibfield  {author} {\bibinfo {author} {\bibfnamefont {T.}~\bibnamefont {Gilbert}},\ }\href {https://doi.org/10.1109/TMAG.2004.836740} {\bibfield  {journal} {\bibinfo  {journal} {IEEE Trans. Magn.}\ }\textbf {\bibinfo {volume} {40}},\ \bibinfo {pages} {3443} (\bibinfo {year} {2004})}\BibitemShut {NoStop}%
\bibitem [{\citenamefont {Vansteenkiste}\ \emph {et~al.}(2014)\citenamefont {Vansteenkiste}, \citenamefont {Leliaert}, \citenamefont {Dvornik}, \citenamefont {Helsen}, \citenamefont {{Garcia-Sanchez}},\ and\ \citenamefont {Van~Waeyenberge}}]{Vansteenkiste2014}%
  \BibitemOpen
  \bibfield  {author} {\bibinfo {author} {\bibfnamefont {A.}~\bibnamefont {Vansteenkiste}}, \bibinfo {author} {\bibfnamefont {J.}~\bibnamefont {Leliaert}}, \bibinfo {author} {\bibfnamefont {M.}~\bibnamefont {Dvornik}}, \bibinfo {author} {\bibfnamefont {M.}~\bibnamefont {Helsen}}, \bibinfo {author} {\bibfnamefont {F.}~\bibnamefont {{Garcia-Sanchez}}},\ and\ \bibinfo {author} {\bibfnamefont {B.}~\bibnamefont {Van~Waeyenberge}},\ }\href {https://doi.org/10.1063/1.4899186} {\bibfield  {journal} {\bibinfo  {journal} {AIP Adv.}\ }\textbf {\bibinfo {volume} {4}},\ \bibinfo {pages} {107133} (\bibinfo {year} {2014})}\BibitemShut {NoStop}%
\bibitem [{\citenamefont {Van~der Walt}\ \emph {et~al.}(2014)\citenamefont {Van~der Walt}, \citenamefont {Sch{\"o}nberger}, \citenamefont {Nunez-Iglesias}, \citenamefont {Boulogne}, \citenamefont {Warner}, \citenamefont {Yager}, \citenamefont {Gouillart},\ and\ \citenamefont {Yu}}]{skimage}%
  \BibitemOpen
  \bibfield  {author} {\bibinfo {author} {\bibfnamefont {S.}~\bibnamefont {Van~der Walt}}, \bibinfo {author} {\bibfnamefont {J.~L.}\ \bibnamefont {Sch{\"o}nberger}}, \bibinfo {author} {\bibfnamefont {J.}~\bibnamefont {Nunez-Iglesias}}, \bibinfo {author} {\bibfnamefont {F.}~\bibnamefont {Boulogne}}, \bibinfo {author} {\bibfnamefont {J.~D.}\ \bibnamefont {Warner}}, \bibinfo {author} {\bibfnamefont {N.}~\bibnamefont {Yager}}, \bibinfo {author} {\bibfnamefont {E.}~\bibnamefont {Gouillart}},\ and\ \bibinfo {author} {\bibfnamefont {T.}~\bibnamefont {Yu}},\ }\href {https://doi.org/10.7717/peerj.453} {\bibfield  {journal} {\bibinfo  {journal} {PeerJ}\ }\textbf {\bibinfo {volume} {2}},\ \bibinfo {pages} {e453} (\bibinfo {year} {2014})}\BibitemShut {NoStop}%
\bibitem [{\citenamefont {Tretiakov}\ \emph {et~al.}(2008)\citenamefont {Tretiakov}, \citenamefont {Clarke}, \citenamefont {Chern}, \citenamefont {Bazaliy},\ and\ \citenamefont {Tchernyshyov}}]{Tretiakov2008}%
  \BibitemOpen
  \bibfield  {author} {\bibinfo {author} {\bibfnamefont {O.~A.}\ \bibnamefont {Tretiakov}}, \bibinfo {author} {\bibfnamefont {D.}~\bibnamefont {Clarke}}, \bibinfo {author} {\bibfnamefont {G.-W.}\ \bibnamefont {Chern}}, \bibinfo {author} {\bibfnamefont {Y.~B.}\ \bibnamefont {Bazaliy}},\ and\ \bibinfo {author} {\bibfnamefont {O.}~\bibnamefont {Tchernyshyov}},\ }\href {https://doi.org/10.1103/PhysRevLett.100.127204} {\bibfield  {journal} {\bibinfo  {journal} {Phys. Rev. Lett.}\ }\textbf {\bibinfo {volume} {100}},\ \bibinfo {pages} {127204} (\bibinfo {year} {2008})}\BibitemShut {NoStop}%
\bibitem [{\citenamefont {Clarke}\ \emph {et~al.}(2008)\citenamefont {Clarke}, \citenamefont {Tretiakov}, \citenamefont {Chern}, \citenamefont {Bazaliy},\ and\ \citenamefont {Tchernyshyov}}]{Clarke2008}%
  \BibitemOpen
  \bibfield  {author} {\bibinfo {author} {\bibfnamefont {D.~J.}\ \bibnamefont {Clarke}}, \bibinfo {author} {\bibfnamefont {O.~A.}\ \bibnamefont {Tretiakov}}, \bibinfo {author} {\bibfnamefont {G.-W.}\ \bibnamefont {Chern}}, \bibinfo {author} {\bibfnamefont {Y.~B.}\ \bibnamefont {Bazaliy}},\ and\ \bibinfo {author} {\bibfnamefont {O.}~\bibnamefont {Tchernyshyov}},\ }\href {https://doi.org/10.1103/PhysRevB.78.134412} {\bibfield  {journal} {\bibinfo  {journal} {Phys. Rev. B}\ }\textbf {\bibinfo {volume} {78}},\ \bibinfo {pages} {134412} (\bibinfo {year} {2008})}\BibitemShut {NoStop}%
\bibitem [{\citenamefont {Braun}(1994)}]{Braun1994}%
  \BibitemOpen
  \bibfield  {author} {\bibinfo {author} {\bibfnamefont {H.-B.}\ \bibnamefont {Braun}},\ }\href {https://doi.org/10.1103/PhysRevB.50.16485} {\bibfield  {journal} {\bibinfo  {journal} {Phys. Rev. B}\ }\textbf {\bibinfo {volume} {50}},\ \bibinfo {pages} {16485} (\bibinfo {year} {1994})}\BibitemShut {NoStop}%
\bibitem [{\citenamefont {Wang}\ \emph {et~al.}(2018)\citenamefont {Wang}, \citenamefont {Yuan},\ and\ \citenamefont {Wang}}]{Wang2018}%
  \BibitemOpen
  \bibfield  {author} {\bibinfo {author} {\bibfnamefont {X.~S.}\ \bibnamefont {Wang}}, \bibinfo {author} {\bibfnamefont {H.~Y.}\ \bibnamefont {Yuan}},\ and\ \bibinfo {author} {\bibfnamefont {X.~R.}\ \bibnamefont {Wang}},\ }\href {https://doi.org/10.1038/s42005-018-0029-0} {\bibfield  {journal} {\bibinfo  {journal} {Commun. Phys.}\ }\textbf {\bibinfo {volume} {1}},\ \bibinfo {pages} {31} (\bibinfo {year} {2018})}\BibitemShut {NoStop}%
\bibitem [{\citenamefont {Virtanen}\ \emph {et~al.}(2020)\citenamefont {Virtanen}, \citenamefont {Gommers}, \citenamefont {Oliphant}, \citenamefont {Haberland}, \citenamefont {Reddy}, \citenamefont {Cournapeau}, \citenamefont {Burovski}, \citenamefont {Peterson}, \citenamefont {Weckesser}, \citenamefont {Bright}, \citenamefont {{van der Walt}}, \citenamefont {Brett}, \citenamefont {Wilson}, \citenamefont {Millman}, \citenamefont {Mayorov}, \citenamefont {Nelson}, \citenamefont {Jones}, \citenamefont {Kern}, \citenamefont {Larson}, \citenamefont {Carey}, \citenamefont {Polat}, \citenamefont {Feng}, \citenamefont {Moore}, \citenamefont {{VanderPlas}}, \citenamefont {Laxalde}, \citenamefont {Perktold}, \citenamefont {Cimrman}, \citenamefont {Henriksen}, \citenamefont {Quintero}, \citenamefont {Harris}, \citenamefont {Archibald}, \citenamefont {Ribeiro}, \citenamefont {Pedregosa}, \citenamefont {{van Mulbregt}},\ and\ \citenamefont {{SciPy 1.0 Contributors}}}]{SciPy}%
  \BibitemOpen
  \bibfield  {author} {\bibinfo {author} {\bibfnamefont {P.}~\bibnamefont {Virtanen}}, \bibinfo {author} {\bibfnamefont {R.}~\bibnamefont {Gommers}}, \bibinfo {author} {\bibfnamefont {T.~E.}\ \bibnamefont {Oliphant}}, \bibinfo {author} {\bibfnamefont {M.}~\bibnamefont {Haberland}}, \bibinfo {author} {\bibfnamefont {T.}~\bibnamefont {Reddy}}, \bibinfo {author} {\bibfnamefont {D.}~\bibnamefont {Cournapeau}}, \bibinfo {author} {\bibfnamefont {E.}~\bibnamefont {Burovski}}, \bibinfo {author} {\bibfnamefont {P.}~\bibnamefont {Peterson}}, \bibinfo {author} {\bibfnamefont {W.}~\bibnamefont {Weckesser}}, \bibinfo {author} {\bibfnamefont {J.}~\bibnamefont {Bright}}, \bibinfo {author} {\bibfnamefont {S.~J.}\ \bibnamefont {{van der Walt}}}, \bibinfo {author} {\bibfnamefont {M.}~\bibnamefont {Brett}}, \bibinfo {author} {\bibfnamefont {J.}~\bibnamefont {Wilson}}, \bibinfo {author} {\bibfnamefont {K.~J.}\ \bibnamefont {Millman}}, \bibinfo {author} {\bibfnamefont {N.}~\bibnamefont {Mayorov}}, \bibinfo {author} {\bibfnamefont {A.~R.~J.}\ \bibnamefont {Nelson}}, \bibinfo {author} {\bibfnamefont {E.}~\bibnamefont {Jones}}, \bibinfo {author} {\bibfnamefont {R.}~\bibnamefont {Kern}}, \bibinfo {author} {\bibfnamefont {E.}~\bibnamefont {Larson}}, \bibinfo {author} {\bibfnamefont {C.~J.}\ \bibnamefont {Carey}}, \bibinfo {author} {\bibfnamefont {{\.I}.}~\bibnamefont {Polat}}, \bibinfo {author} {\bibfnamefont {Y.}~\bibnamefont {Feng}}, \bibinfo {author} {\bibfnamefont {E.~W.}\ \bibnamefont {Moore}}, \bibinfo {author} {\bibfnamefont {J.}~\bibnamefont {{VanderPlas}}}, \bibinfo {author} {\bibfnamefont {D.}~\bibnamefont {Laxalde}}, \bibinfo {author} {\bibfnamefont {J.}~\bibnamefont {Perktold}}, \bibinfo {author} {\bibfnamefont {R.}~\bibnamefont {Cimrman}}, \bibinfo {author} {\bibfnamefont {I.}~\bibnamefont {Henriksen}}, \bibinfo {author} {\bibfnamefont {E.~A.}\ \bibnamefont {Quintero}}, \bibinfo {author} {\bibfnamefont {C.~R.}\ \bibnamefont {Harris}}, \bibinfo {author} {\bibfnamefont {A.~M.}\
  \bibnamefont {Archibald}}, \bibinfo {author} {\bibfnamefont {A.~H.}\ \bibnamefont {Ribeiro}}, \bibinfo {author} {\bibfnamefont {F.}~\bibnamefont {Pedregosa}}, \bibinfo {author} {\bibfnamefont {P.}~\bibnamefont {{van Mulbregt}}},\ and\ \bibinfo {author} {\bibnamefont {{SciPy 1.0 Contributors}}},\ }\href {https://doi.org/10.1038/s41592-019-0686-2} {\bibfield  {journal} {\bibinfo  {journal} {Nat. Methods}\ }\textbf {\bibinfo {volume} {17}},\ \bibinfo {pages} {261} (\bibinfo {year} {2020})}\BibitemShut {NoStop}%
\end{thebibliography}%

\appendix

\section{Goldstone mode limit}

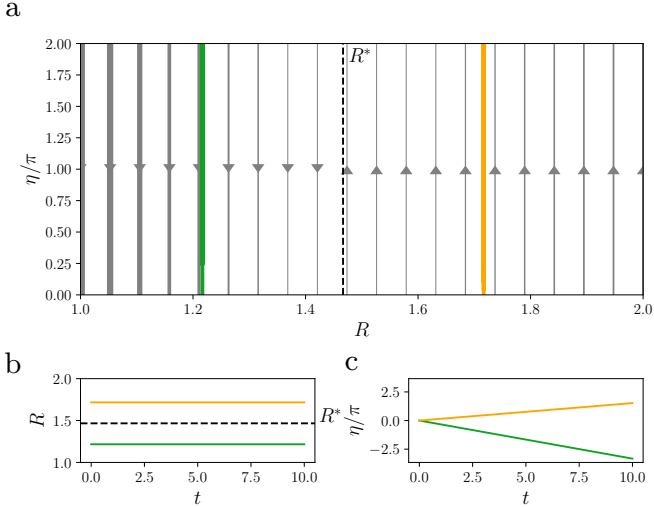
\begin{figure}[h!]
    \centering
    \begin{tikzpicture}    
        \node {\includegraphics[width=\linewidth]{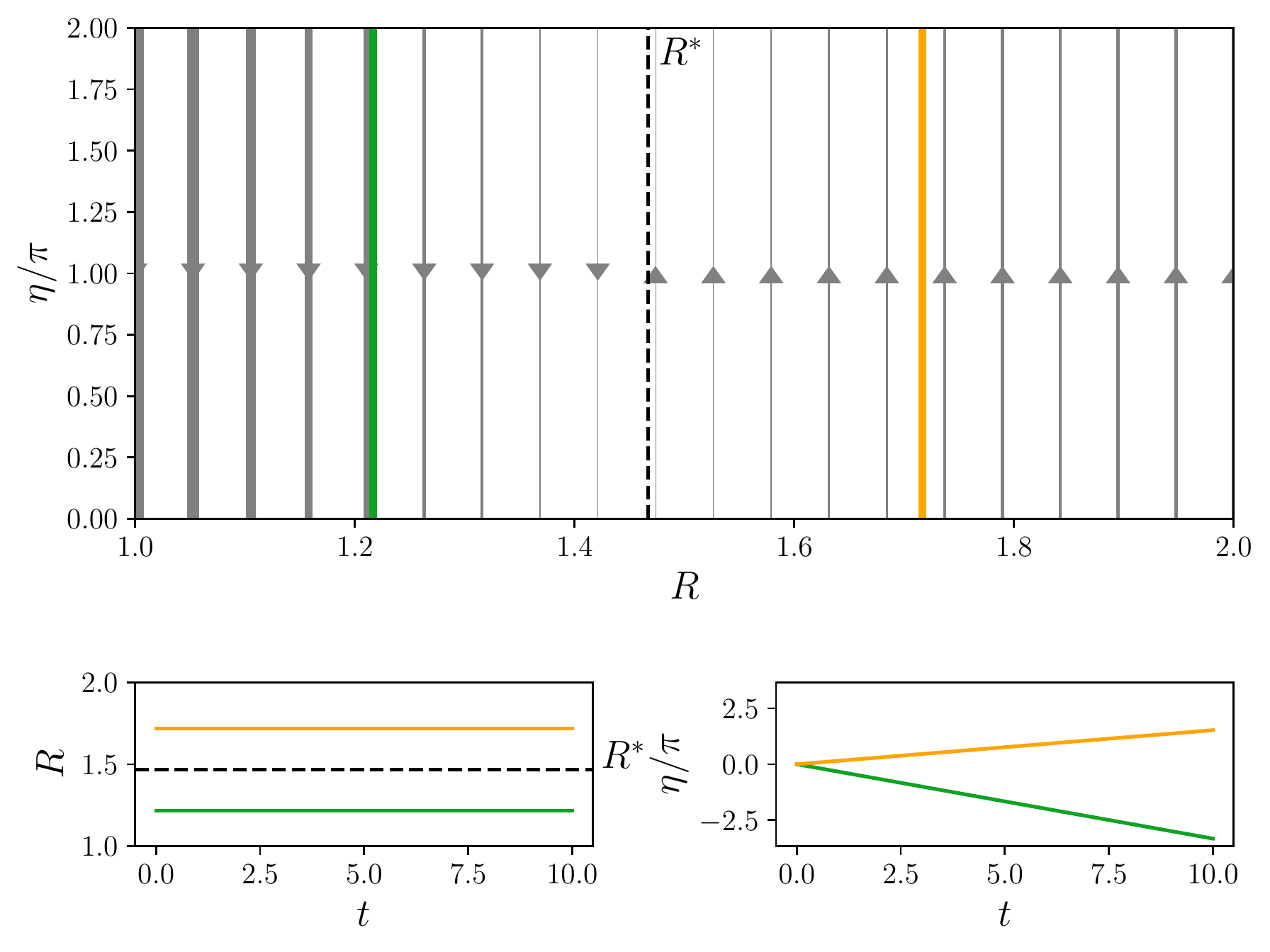}};
        \node[scale=1.25] at (-4.3, 3.5) {a};
        \node[scale=1.25] at (-4.3, -1.2) {b};
        \node[scale=1.25] at (0.2, -1.2) {c};
    \end{tikzpicture}
    \caption{
    Collective coordinate dynamics in the limit $E_0 = \alpha = 0$.
    a) Phase portrait of Thiele equation solutions.
    In this limit all the trajectories are parallel to the $\eta$ axis and, given the helicity’s $S^1$ topology, they describe helicity rotations at a constant radius.
    The arrows indicate the direction of helicity rotation (clockwise for $R<R^*$ and counterclockwise for $R>R^*$) while the trajectories’ thickness is proportional to the value of $\dot{\eta}$, which vanishes at $R = R^*$.
    For two selected trajectories ($R = R^* - 0.25$ in green and $R = R^* + 0.25$ in orange) we plot the time evolution of their radius b) and helicity c).
    }
    \label{fig:PhasePortraitNoDamping}
\end{figure}

Figure~\ref{fig:PhasePortraitNoDamping} summarizes the solutions of the Thiele equations in the limit of zero damping and absence of an externally applied electric field, i.e.\ $E_0 = \alpha = 0$. In this limit, the Thiele equations decouple and energy is conserved. Therefore the radius remains constant for any initial skyrmion ansatz. For the critical radius of $R=R^*$, the skyrmion has its minimal energy.
In contrast, the helicity is a Goldstone mode and $\dot{\eta}$ depends only on the skyrmion radius.
Moreover, $\dot{\eta}$ swaps sign at $R=R^*$, where for $R<R^*$ ($R>R^*$) the skyrmion's helicity rotation evolves monotonically clockwise (counterclockwise).

\section{Collective coordinate modeling: radial and energy results}

\begin{figure}[H]
	\centering
        \begin{tikzpicture}
            \node {\includegraphics[width=\linewidth]{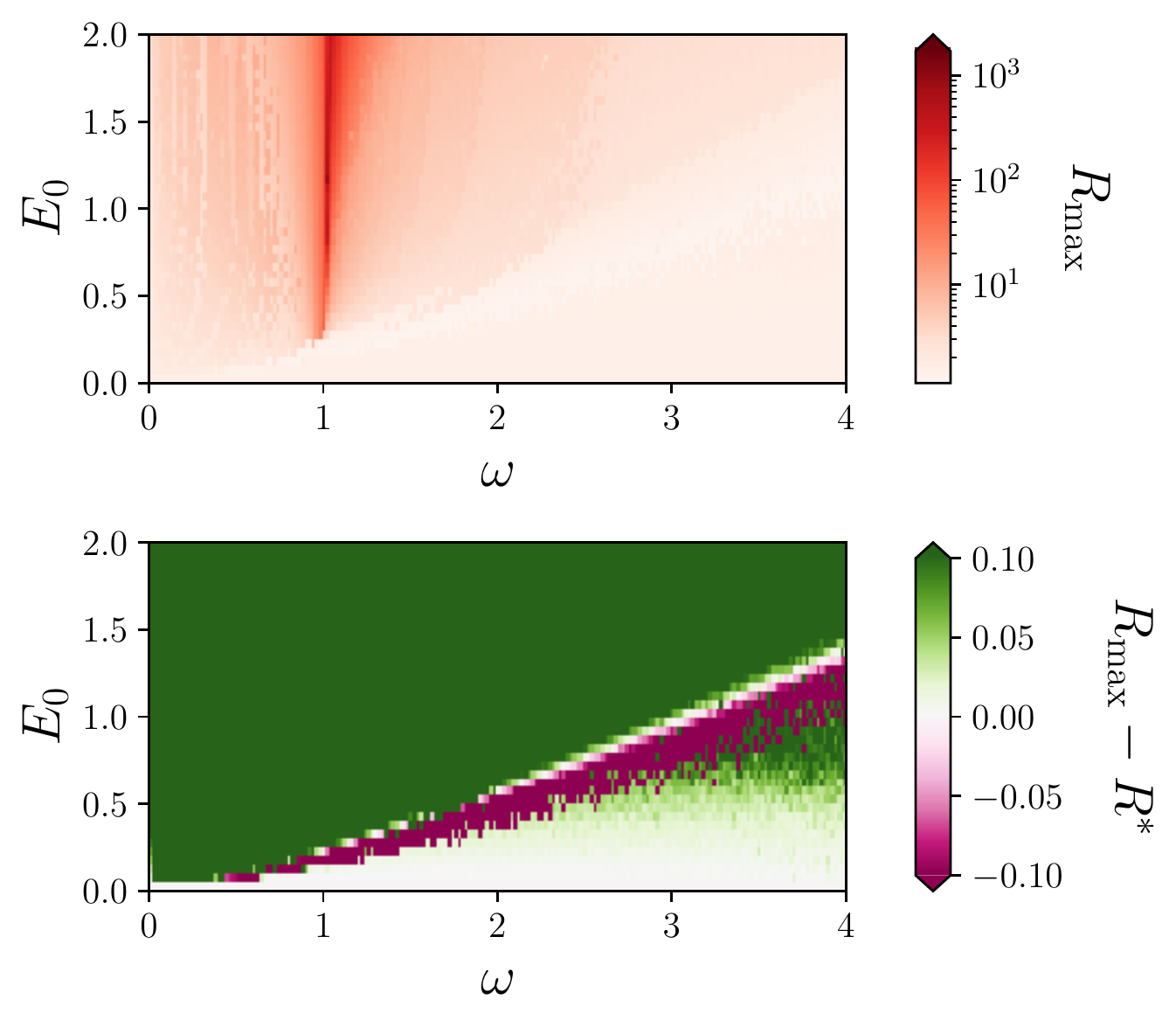}};
            \node[scale=1.25] at (-4.0, 3.8) {a};
            \node[scale=1.25] at (-4.0, 0.2) {b};
        \end{tikzpicture}
	\caption{a) Maximal skyrmion radius $R_{\rm{max}}$ and b) difference between the maximal and the relaxed skyrmion radii $R_{\rm{max}} - R^*$, as functions of $E_0$ and $\omega$ obtained from Thiele equations solutions.
            Note that the plotted range of $R_{\rm{max}} - R^*$ is restricted to $[-0.1, 0.1]$, thus the divergence of $R_{\rm{max}}$ at the Kittel resonance frequency seen in a) is not shown.}
\label{fig:RadiusPhaseDiagrams}
\end{figure}

In the phase diagram shown in Fig.~\ref{fig:PhaseDiagram}a of the main text, the regions of monotonic (dark pink) and non-monotonic (light pink) helicity evolution are separated by a straight line. 
To elucidate its meaning, in Fig.~\ref{fig:RadiusPhaseDiagrams}a we show $R_{\rm{max}}$, the maximum value attained by the radius in the long-time limit of the Thiele equations solution, as a function of $E_0$ and $\omega$.
Since $R_{\rm{max}}$ becomes large in the Kittel resonance region and makes it hard to observe features elsewhere, in Fig.~\ref{fig:RadiusPhaseDiagrams}b we plot $R_{\rm{max}} - R^*$ ($R^*$ is the relaxed skyrmion radius in the absence of electric field) and restrict its range of values to $[-0.1, 0.1]$ to enhance features away from resonance.
The white straight line in Fig.~\ref{fig:RadiusPhaseDiagrams}b corresponds to solutions with $R_{\rm{max}} \approx R^*$ and coincides with the straight line in Fig.~\ref{fig:PhaseDiagram}a.
Note that $R_{\rm{max}} > R^*$ above the white straight line.
Additionally, our analysis of the Thiele equations shows that when $R > R^*$ ($R < R^*$) the helicity increases (decreases).
Therefore, for values of $E_{0}$ and $\omega$ in the phase diagram's light pink region, the skyrmion radius crosses $R = R^*$ causing the helicity to reverse direction, thus becoming non-monotonic.

\begin{figure}[H]	\label{fig:EnergyPhaseDiagram}
	\centering
\includegraphics[width=\linewidth]{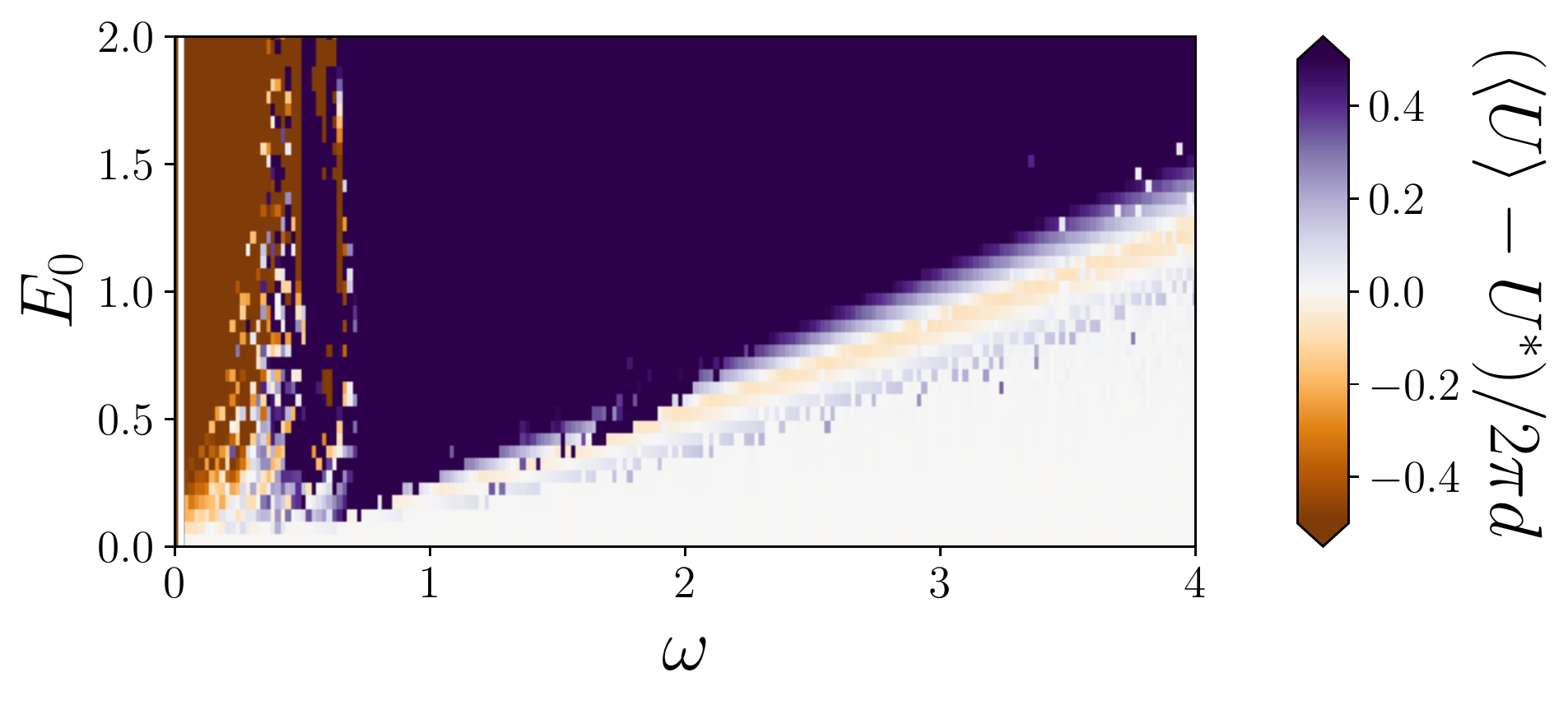}
	\caption{
    Time-average of the energy of a driven skyrmion $\langle U \rangle$ over one period of the driving electric field.
    The energy of a relaxed skyrmion with no electric field $U^*$ has been subtracted.
    For values of $E_0$ and $\omega$ where $\langle U \rangle = U^*$ the energy injected by the driving AC electric field is on average balanced out by Gilbert damping dissipation.
    }
	\label{fig:EnergyMinusSkyrmion}
\end{figure}

The construction of a spacetime magnetic hopfion hinges upon the skyrmion dynamics settling into a time-periodic steady state.
In our construction route 1, we activate the helicity rotation by an AC electric field that injects energy while being counteracted by Gilbert damping dissipation.
Figure~\ref{fig:EnergyMinusSkyrmion} shows the time-average of the energy of a driven skyrmion $\langle U \rangle$ with the energy of a relaxed skyrmion in the absence of the electric field $U^*$ subtracted.
Upon comparison with the phase diagram in Fig.~\ref{fig:PhaseDiagram}a, we conclude that the sought-after time-periodic steady state where the helicity rotates establishes for values of $E_0$ and $\omega$ where $\langle U \rangle \approx U^*$, namely, where energy injection and dissipation balance each other out on average.

\end{document}